%% file: main.tex
\documentclass[acmlarge, nonacm]{acmart}

\AtBeginDocument{%
  }

\usepackage{xspace}
\usepackage{caption}
\usepackage{subcaption}
\usepackage{url}
\usepackage[ruled,vlined,linesnumbered]{algorithm2e}
\usepackage{algpseudocode}
\usepackage{bm}
\usepackage{amsmath}
\usepackage{color}
\usepackage{soul}
\usepackage{array}
\usepackage{gensymb}
\usepackage{tabularx}
\usepackage{soul}
\usepackage{float}
\usepackage{booktabs}
\usepackage{multirow}
\usepackage{tablefootnote}

\graphicspath{{../}{figures/}}
\newcommand{\ie}{{\it i.e.}\xspace}

\newcommand{\etal}{{\it et~al.}\xspace}

\newcommand{\systemname}{\textit{MobileAF}\xspace}

\setcopyright{acmlicensed}
\copyrightyear{2018}
\acmYear{2018}
\acmDOI{XXXXXXX.XXXXXXX}

\acmConference[Conference acronym 'XX]{Make sure to enter the correct
  conference title from your rights confirmation emai}{June 03--05,
  2018}{Woodstock, NY}
\acmISBN{978-1-4503-XXXX-X/18/06}

\begin{document}

\title{Atrial Fibrillation Detection System via Acoustic Sensing for Mobile Phones}

\author{Xuanyu Liu}
\orcid{0009-0001-2825-311X}
\affiliation{%
  \institution{Research Institute of Trustworthy Autonomous Systems, Department of Computer Science and Engineering, Southern University of Science and Technology}
  \city{Shenzhen}
  \country{China}
}

\author{Jiao Li}
\orcid{0000-0002-3918-4922}
\affiliation{%
  \institution{Research Institute of Trustworthy Autonomous Systems, Department of Computer Science and Engineering, Southern University of Science and Technology}
  \city{Shenzhen}
  \country{China}
}

\author{Haoxian Liu}
\orcid{0009-0004-7550-1325}
\affiliation{%
  \institution{Research Institute of Trustworthy Autonomous Systems, Department of Computer Science and Engineering, Southern University of Science and Technology}
  \city{Shenzhen}
  \country{China}
}

\author{Zongqi Yang}
\orcid{0009-0006-4191-8290}
\affiliation{%
  \institution{Research Institute of Trustworthy Autonomous Systems, Department of Computer Science and Engineering, Southern University of Science and Technology}
  \city{Shenzhen}
  \country{China}
}

\author{Yi Huang}
\orcid{0009-0008-0039-4026}
\affiliation{
  \institution{Department of Cardiology, Southern University of Science and Technology Hospital}
  \city{Shenzhen}
  \country{China}
}

\author{Jin Zhang}
\email{zhangj4@sustech.edu.cn}
\orcid{0000-0002-2674-0918}
\affiliation{%
  \institution{Research Institute of Trustworthy Autonomous Systems, Department of Computer Science and Engineering, Southern University of Science and Technology}
  \city{Shenzhen}
  \country{China}
}
\authornote{Corresponding author}

\renewcommand{\shortauthors}{Liu et al.}

\begin{abstract}
Atrial fibrillation (AF) is characterized by irregular electrical impulses originating in the atria, which can lead to severe complications and even death. Due to the intermittent nature of the AF, early and timely monitoring of AF is critical for patients to prevent further exacerbation of the condition. Although ambulatory ECG Holter monitors provide accurate monitoring, the high cost of these devices hinders their wider adoption. Current mobile-based AF detection systems offer a portable solution, however, these systems have various applicability issues such as being easily affected by environmental factors and requiring significant user effort. To overcome the above limitations, we present \systemname, a novel smartphone-based AF detection system using speakers and microphones. In order to capture minute cardiac activities, we propose a multi-channel pulse wave probing method. In addition, we enhance the signal quality by introducing a three-stage pulse wave purification pipeline. What's more, a ResNet-based network model is built to implement accurate and reliable AF detection. We collect data from 23 participants utilizing our data collection application on the smartphone. Extensive experimental results demonstrate the superior performance of our system, with 97.9\% accuracy, 96.8\% precision, 97.2\% recall, 98.3\% specificity, and 97.0\% F1 score.
\end{abstract}



\maketitle

\input{Body/intro}
\input{Body/feasibility}

\input{Body/overview}
\input{Body/pulse_wave_probing}   
\input{Body/pulse_wave_extraction}  
\input{Body/pulse_wave_quality_filter}
\input{Body/pulse_wave_purification}

\input{Body/AF_detection}

\input{Body/implementation}
\input{Body/evaluation}

\input{Body/related}

\input{Body/discussion}
\input{Body/conclusion}



\bibliographystyle{ACM-Reference-Format}
\bibliography{ref}

\end{document}

%% file: Body/intro.tex
\section{Introduction}
\label{s:intro}
Atrial fibrillation (AF), characterized by irregular electrical impulses in the atria, is the most prevalent clinically significant cardiac rhythm disorder in the 21st century \cite{AF_significance}. Over the last two decades, the incidence of AF has risen substantially, with forecasts predicting a further increase over the next 30 years, posing a significant public health challenge worldwide \cite{AF-Intro}. According to The Global Burden of Disease 2019 study, nearly 59 million people were living with AF in 2019. This disorder not only increases the risk of ischemic stroke but is also associated with other severe outcomes like heart failure and additional cardiovascular complications \cite{AF-With-Stroke}. Given its growing prevalence and the potential for serious health consequences, the early detection of AF is critical.

However, diagnosing AF presents challenges in practice. Firstly, a considerable number of individuals with AF are asymptomatic, as evidenced by at least one-third of patients experiencing no symptoms \cite{silentAF}. This lack of symptoms discourages them from seeking cardiovascular examinations at hospitals. Additionally, AF tends to be intermittent, further complicating diagnosis. Furthermore, the clinical diagnosis of AF heavily relies on electrocardiogram (ECG), which demands costly medical facilities for data collection and specialized medical expertise for interpretation. 

To tackle the aforementioned issues, many research efforts have focused on utilizing mobile devices, such as smartwatches and smartphones, for AF detection. A common approach in current studies involves using smartwatches equipped with photoplethysmography (PPG) sensors to gather pulse wave data for analysis \cite{EarlyDetectionAf, WristPPG-1, WristPPG-2, WristPPG-3, WristPPG-4, WristPPG-5, WristPPG-6}. To mitigate the influence of motion artifacts, some works utilize inertial measurement unit (IMU) data to assess data quality or to denoise \cite{WristPPGIMU-1, WristPPGIMU-2, WristPPGIMU-3, WristPPGIMU-4, WristPPGIMU-5}. Furthermore, a lot of studies use smartphone cameras and flashlights to replace PPG sensors to obtain PPG data and detect AF \cite{CameraPPG-1, CameraPPG-2, CameraPPG-3, CameraPPG-4, CameraPPG-5, CameraPPG-6, CameraPPG-7, CameraPPG-8, CameraPPG-9}. Although PPG-based approaches are of relatively low cost, they are sensitive to skin tone \cite{PPGSkinTone} and may disturb users with emitted light, especially when used at night. Besides, some flashlights will produce heat for long-term use, diminishing user experience. Another innovative method involves analyzing seismocardiography (SCG) or ballistocardiography (BCG) data captured by the IMU of smartphones \cite{SCGAF, SCG1, SCG2, SCG3}. IMU sensors are common on smartphones, making this approach feasible. However, it requires users to lie down and place smartphones on their chests, demanding significant user effort and applicable only in limited situations. The third popular approach involves using mobile ECG devices to obtain heart activity data for analysis \cite{ECG2, ECG3, MobileECG-1, MobileECG-2, MobileECG-3, MobileECG-4, MobileECG-5, MobileECG-6}. Although ECG serves as the gold standard for AF detection, mobile ECG devices cannot provide data of the same quality as medical ECG devices, leading to diminished accuracy. What's more, this approach typically has a high cost and still requires additional devices, limiting their widespread use.

\begin{figure}
  \centering
  \includegraphics[width=5in]{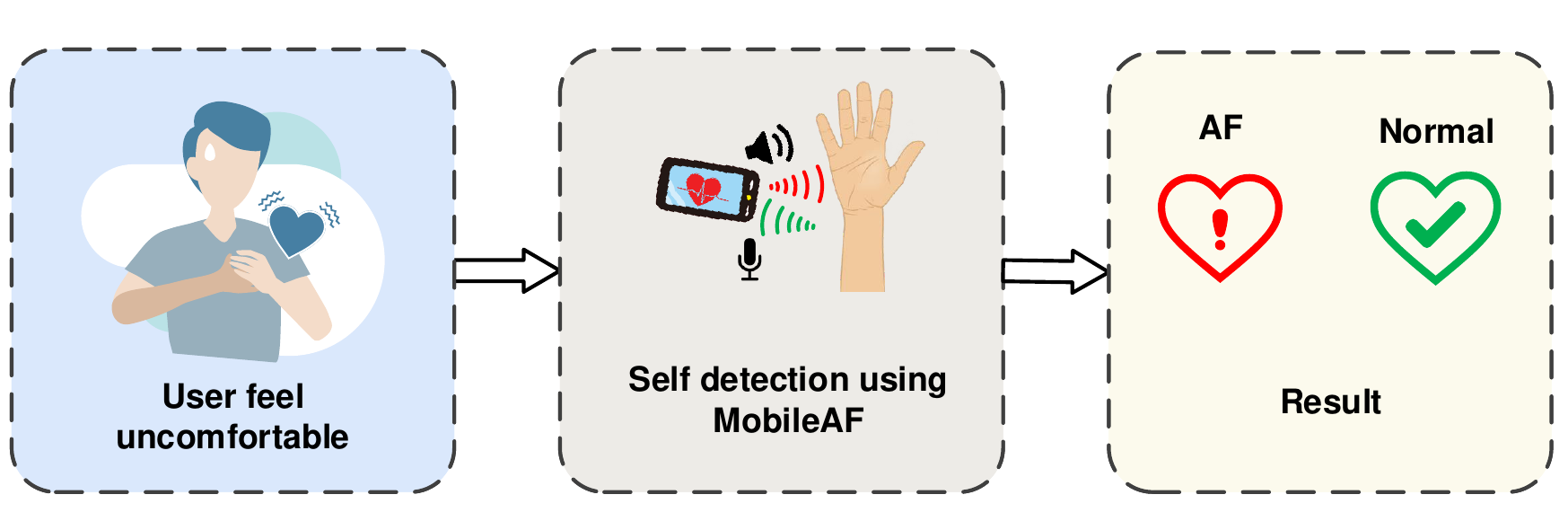}
  \caption{\systemname application scenario.}
  \vspace{-0.2in}
  \label{f: scenario}
\end{figure}

To solve these problems, we propose \systemname, a system that leverages the built-in microphones and speakers of commercial off-the-shelf (COTS) smartphones for AF detection. Fig. \ref{f: scenario} shows a potential application scenario of our system. \systemname offers a user-friendly solution to AF screening. When a user experiences discomfort, they can simply use their smartphone to perform a self-test anywhere. This eliminates the need for additional equipment and provides immediate results, addressing the issue of delayed diagnosis and treatment due to intermittent episodes of AF. The system can be particularly helpful in AF screening because users may fail to be detected with AF when they visit the hospital, as many AF cases are intermittent. 

Although promising, three challenges underlie the design:

\begin{itemize}
    \item The feasibility of using acoustic sensing to capture pulse waves on wrists via smartphones remains unexplored. While APG \cite{acous4} demonstrates the potential of using in-ear microphones to detect pulse waves, it leverages the unique structure of the ear canal, which differs significantly from the wrist. Therefore, we conducted a feasibility study and confirmed that it is possible to use acoustic sensing technologies to extract pulse waves from the wrist.

    \item The acquisition of pulse waves depends heavily on the frequency of the probing signal and the position of the device, presenting critical challenges for system robustness. To address these issues, we develop a multi-channel probing signal that reduces the impact of the frequency selection problem. Moreover, we introduce a pulse wave quality assessment module to filter out poor-quality data and provide guidance on positioning devices, ensuring the system's reliable results.
    
    \item The acquired pulse waves may be distorted by both the transmission through the device's internal structure and the user's motion artifacts, which complicates the precise extraction of pulse waves. We propose denoising algorithms that effectively eliminate static noise components to mitigate the influence of internal structural transmission. Additionally, we use stationary wavelet transformation (SWT) to reconstruct signals, reducing the effects of motion artifacts.
\end{itemize}

In this study, we present the following contributions:	
\begin{itemize}
    \item To the best of our knowledge, \systemname is the first AF detection system implemented on COTS smartphones using acoustic sensing technology. This innovative approach utilizes smartphones' built-in microphones and speakers for non-intrusive pulse wave monitoring, eliminating the need for additional sensors.
    \item We present a fine-grained AF detection pipeline. Initially, a multi-channel pulse wave probing signal and an associated extraction algorithm are implemented to capture subtle pulse wave variations. Subsequently, a pulse wave quality assessment algorithm is introduced to aid device positioning and exclude severely distorted signals. The filtered data are then further purified by eliminating static components and removing motion artifacts. Finally, the purified data are fed into a ResNet-based AF detector. This pipeline is designed to deliver precise and reliable AF detection results across diverse scenarios.
    \item We conduct extensive experiments involving 23 subjects, comprising 9 individuals with AF and 14 healthy individuals under various scenarios. The results demonstrate the high performance of our system, achieving an overall accuracy of 97.9\%, recall of 97.2\%, precision of 96.8\%, specificity of 98.3\%, and an F1 score of 97.0\%.
\end{itemize}

%% file: Body/feasibility.tex
\begin{figure}
  \centering
  \includegraphics[width=5in]{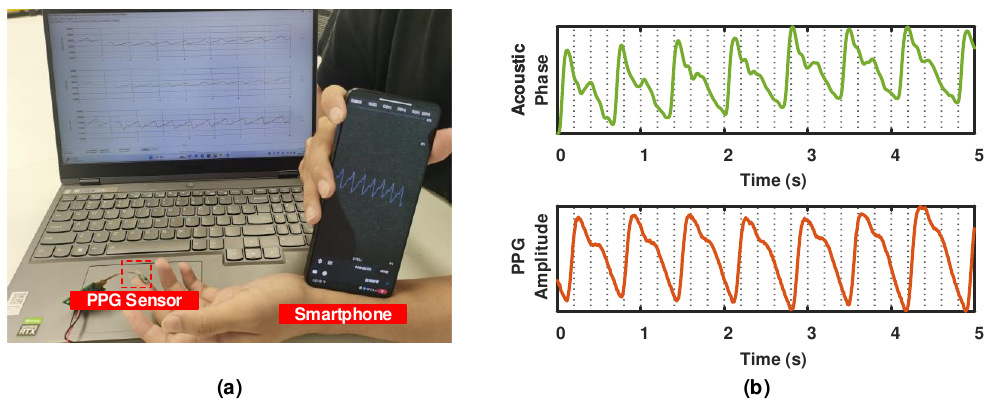}
  \caption{Feasibility study setup and result. \textbf{(a)}: Scenario of feasibility study. The PPG sensor is placed at the fingertip, and the smartphone is placed at the wrist above the radial artery. \textbf{(b)}: Comparison of PPG and acoustic phase changes. The phase waveform shows a high similarity with the PPG waveform.}
  \vspace{-0.2in}
  \label{f: feasibility}
\end{figure}

\section{Feasibility Study}
\label{s: feasibility_study}

Reliable detection of atrial fibrillation requires the acquisition of high-fidelity pulse waves. The radial artery, located on the thumb side of the wrist, has strong pulsations, and these pulsations due to cardiac activity can be sensed through the fingertip by pressing on it with the finger. \textit{Similarly, if the speaker and microphone of a smartphone are pressed on the radial artery like a finger, can the tiny deformations of the skin surface caused by the heartbeat be sensed?} 

To answer this question, we conduct a feasibility study. In particular, to analyze the small movements caused by the wrist's radial artery, we propose to monitor phase changes based on I/Q demodulation. The study aims to compare the correlation between acoustic phase changes and PPG waveform. A Redmi Note 13 Pro smartphone and a MAXM86161 PPG sensor are utilized to concurrently acquire acoustic and PPG data, respectively, as illustrated in Fig.~\ref{f: feasibility}(a). The smartphone, positioned above the wrist with its microphone facing the radial artery, emits an 18,000 Hz sinusoidal signal through the speaker and records the received signal for analysis. Subsequently, the phase changes are obtained using the algorithm detailed in Sec.~\ref{s: pulse_wave_extraction}. At the same time, the PPG sensor placed at the fingertip is recording PPG data as a comparison.

The experimental results, presented in Fig.~\ref{f: feasibility}(b), exhibit a strong correlation between the phase changes and PPG data, demonstrating the feasibility of extracting high-fidelity pulse waves from the wrist using smartphone speakers and microphones. This opens up the potential for detecting AF using acoustic sensing.

%% file: Body/overview.tex
\section{Overview}
\label{s: overview}

\systemname probes the pulse waves from the vibrations at the wrist to enable AF detection. However, probing pulse waves is challenging because the vibrations are minute. Thus, we need to design a signal processing approach with high sensitivity. Additionally, the amplitudes of vibrations vary across the wrist's positions, requiring us to design a data quality assessment approach to filter out low-quality data and provide guidance on device positioning. Furthermore, the probing signals are easily distorted by both the transmission through the device's internal structure and the user's motion artifacts, so we need to design algorithms to eliminate static components caused by the inner structural transmission and remove motion artifacts.

\begin{figure}[ht]
  \centering
  \includegraphics[width=5.5in]{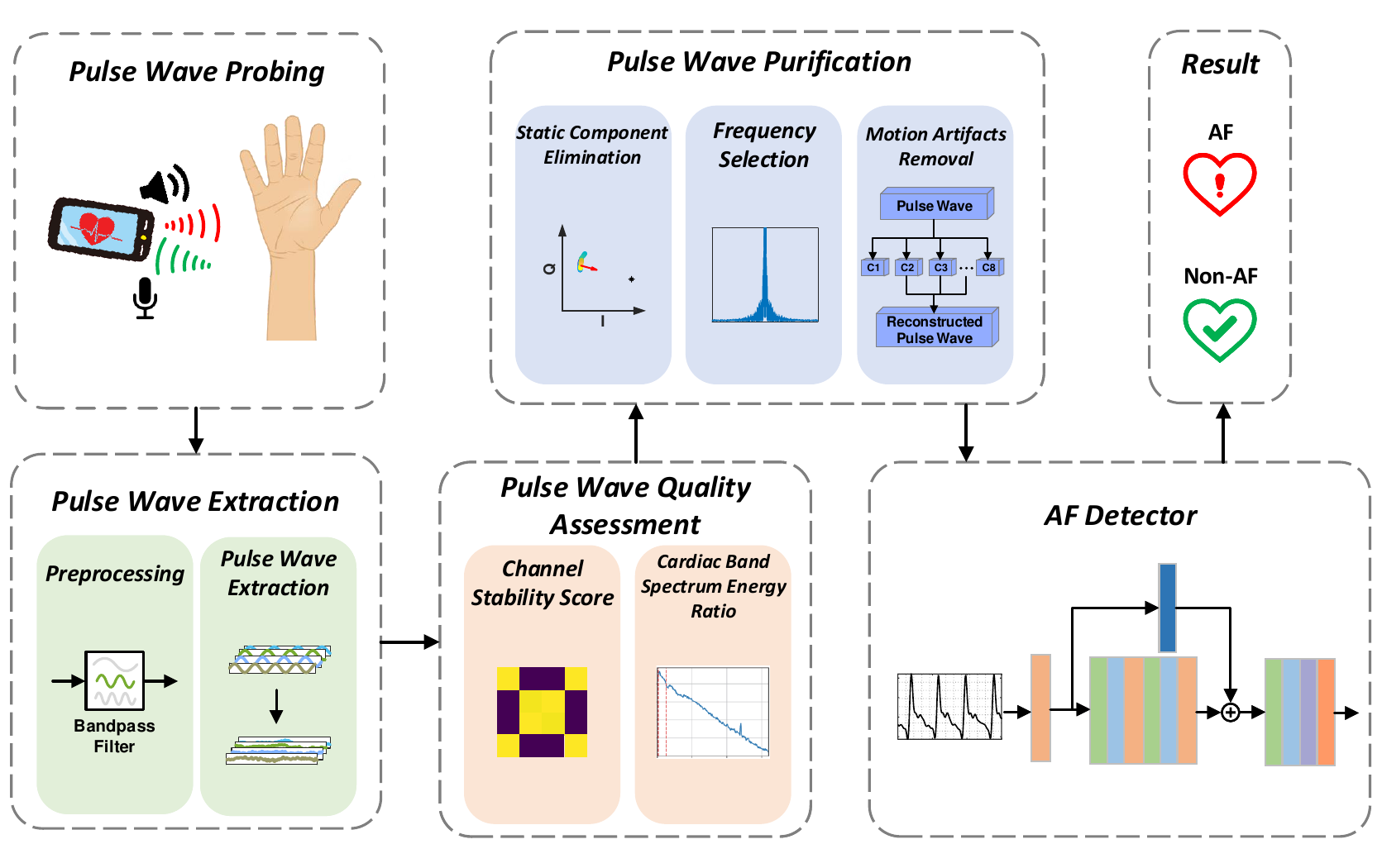}
  \caption{\systemname system overview.}
  \vspace{-0.2in}
  \label{f: overview}
\end{figure}

In response to these demands, we designed \systemname as shown in Fig.~\ref{f: overview}, which consists of five modules: 
\textbf{(i) Pulse Wave Probing.} To reduce the impact of the frequency selection problem, \systemname emits probing signals of multi-frequency sine waves to capture minute pulse waves.
\textbf{(ii) Pulse Wave Extraction.} To remove noise from daily life, the received signals are filtered using a bandpass filter. Subsequently, pulse waves are extracted from the signal phase changes.
\textbf{(iii) Pulse Wave Quality Assessment.} Based on channel stability and spectrum, two metrics are designed to filter out low-quality data that is severely distorted or interfered with and provide guidance on positioning device.
\textbf{(iv) Pulse Wave Purification.} To further correct the distorted pulse waves, we propose a PCA-based algorithm to eliminate the static components. Then, to reduce the computational overhead for subsequent modules, two analytic metrics are proposed to select pulse waves from the best frequency. High-quality pulse waves are further purified with an SWT-based interference removal module to remove motion artifacts.
\textbf{(v) AF Detector.} The purified pulse waves are fed into a ResNet-based AF detector to conduct final AF detection.

%% file: Body/pulse_wave_probing.tex
\section{Pulse Wave Probing}
\label{s: pulse_wave_probing}

To detect AF, it is first necessary to obtain a pulse wave signal that is highly correlated with cardiac activity, \ie, we should capture the rhythmic changes in the surface of the radial artery at the wrist caused by heartbeats. To capture pulse waves, an intuitive approach is to transmit a single-channel probing signal, which only consists of sinusoidal waves of a specific frequency, and record the phase changes between the transmitted signal and the received signal. However, single-channel probing signals do not guarantee reliable pulse wave acquisition because the acoustic transmission from the speaker to the microphone is always complex and is likely to have frequency-selective attenuation. Specifically, we observed that pulse wave varies significantly across individuals at different carrier frequencies. To address this issue, we design a multi-channel detection signal, whereby the signal consists of multiple sine waves as independent carriers ranging from 18 kHz to 21 kHz in 1 kHz step increments. In this way, pulse waves can always be captured by one or multiple channels.

%% file: Body/pulse_wave_extraction.tex
\section{Pulse Wave Extraction}
\label{s: pulse_wave_extraction}
In this section, we delve into the methodologies employed to perform pulse wave extraction, laying the groundwork for effective AF detection.

\subsection{Pre-processing}
\label{ss: received_signal_pre-processing}
Upon reception by the microphone, the signal may be contaminated with ambient noise. Also, channels may interfere with each other in the following processing. To address these issues, a bandpass Butterworth filter is applied to each channel. The filter is configured with a lower cutoff frequency of $f_{c_i} - 50$ Hz and an upper cutoff frequency of $f_{c_i} + 50$~Hz, where $f_{c_i}$ is the carrier frequency of the i-th channel. This step ensures a clean signal for subsequent analysis.

\subsection{Pulse Wave Extraction}
\label{ss: pulse_wave_ext }
For detecting AF, \systemname needs to extract the pulse waves in the received signals, which reflect cardiac activities. To achieve this, we calculate the phases of the four channels. The acoustic signal transmitted from a smartphone speaker can be represented as $S(t) = \sum_{i=1}^{n} \alpha_{i}cos(2\pi f_{c_i}t)$, where $\alpha_i$ is the gain coefficients for channels due to the speaker's non-uniform frequency response, and $n=4$. Then, the signal transmits through the wrist skin, arrives at the smartphone microphone, and gets pre-processed by the bandpass filter. The received signal after pre-processing regarding the carrier frequency $f_{c_i}$ can be denoted as: 
\begin{align*}
R_{i}(t) &= A_{i}(t) \cos(2\pi f_{c_i} t - \theta_{c_i}(t) - \theta_{p_i}),
\end{align*}
where $A_{i}(t)$ is the amplitude of the received signal, $\theta_{c_i}(t)$ represents the phase of the channel, and $\theta_{p_i}$ denotes the phase offset due to hardware delay and system noise. In most cases, $\theta_{p_i}$ can be considered as a constant and does not change over a short time.

Next, we utilize In-phase and Quadrature (I/Q) demodulation to calculate the phase. We first multiply the received signal $R_{i}(t)$ by $\cos(2\pi f_{c_i} t)$ to give a signal with the addition of one low-frequency component and multiple high-frequency components, which can be expressed as:
\begin{equation}
\begin{aligned}
    R_{i}(t)\cos(2\pi f_{c_i} t) 
    &= A_{i}(t) \cos(2\pi f_{c_i} t - \theta_{c_i}(t) - \theta_{p_i})\cos(2\pi f_{c_i} t) \\
    &= \frac{1}{2}A_{i}(t)[\cos(\theta_{c_i}(t) + \theta_{p_i}) + \cos(4\pi f_{c_i} t - \theta_{c_i}(t) - \theta_{p_i})].
\end{aligned}
\label{eq:1}
\end{equation}
In this equation, $\cos(4\pi f_{c_i} t - \theta_{c_i}(t) - \theta_{pi})$ is naturally of high-frequency component, and $\cos(\theta_{c_i}(t) + \theta_{p_i})$ is the low-frequency component. After processing the signal with a low-pass filter, we can obtain the low-frequency component, which is called $I$ signal:
\begin{eqnarray}
I_i(t) = \frac{1}{2}A_{i}(t)\cos(\theta_{c_i}(t) + \theta_{p_i}).
\end{eqnarray}
Similarly multiplying the received signal $R_{i}(t)$ by $\sin(2\pi f_{c_i} t)$ and passing it through a low-pass filter yields the $Q$ signal:\\
\begin{eqnarray}
Q_i(t) = \frac{1}{2}A_{i}(t)\sin(\theta_{c_i}(t) + \theta_{p_i}).
\end{eqnarray}

Once $I_i$ and $Q_i$ signals are obtained, they are downsampled to 128 Hz to accelerate the computation for subsequent procedures. To obtain the phase changes, \ie, the extracted pulse wave $\phi_{est}^{i}$ for the i-th channel, we only need to calculate the $arctan(\frac{Q}{I})$:
\begin{align}
\label{e: final_CPR}
\phi_{est}^{i}(t) &= \arctan\left(\frac{Q}{I}\right) \nonumber \\
           &= \arctan\left(\frac{\frac{1}{2}A_{i}(t)\sin(\theta_{c_i}(t) + \theta_{p_i})}{\frac{1}{2}A_{i}(t)\cos(\theta_{c_i}(t) + \theta_{p_i})}\right) \nonumber \\
           &= \theta_{c_i}(t) + \theta_{p_i} .
\end{align}

\begin{figure}[ht]
  \centering
  \includegraphics[width=\linewidth]{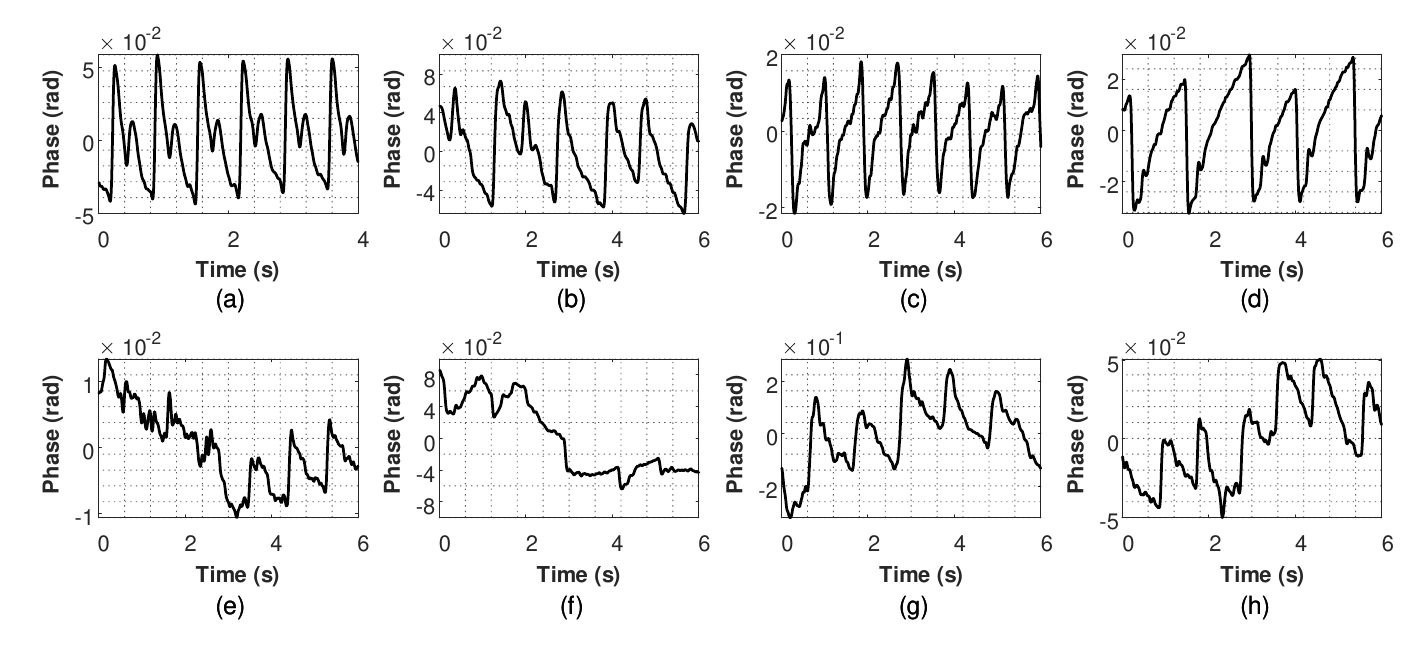}
  \caption{Comparison of extracted pulse waves between subjects with AF and NSR, illustrating various patterns. Results labeled \textbf{(a)}, \textbf{(c)}, \textbf{(e)}, and \textbf{(f)} pertain to NSR subjects, while \textbf{(b)}, \textbf{(d)}, \textbf{(f)}, and \textbf{(g)} correspond to AF subjects. Subfigures \textbf{(a)} and \textbf{(b)} depict ideal cases, \textbf{(c)} and \textbf{(d)} represent reversed cases, \textbf{(e)} and \textbf{(f)} show distorted cases, \textbf{(g)} and \textbf{(h)} illustrate baseline-drift cases.}
  \label{f: CPB_results}
\end{figure}

Fig.~\ref{f: CPB_results} illustrates the distinct patterns of extracted pulse waves in subjects without AF and those with AF. High-quality signals like Fig.~\ref{f: CPB_results}(a) and Fig.~\ref{f: CPB_results}(b) exhibit a clear and stable heartbeat pattern, where AF subjects demonstrate markedly different features than normal sinus rhythm (NSR) subjects. However, some signals depicted in Fig.~\ref{f: CPB_results}(c) through Fig.~\ref{f: CPB_results}(h) exhibit inverted, distorted, or baseline-drift patterns, which further complicates AF detection. These unexpected patterns present a significant challenge for the design of \systemname, which will be addressed in the following sections.

%% file: Body/pulse_wave_quality_filter.tex
\section{Pulse Wave Quality Assessment}
\label{s: pulse_wave_quality_screening}

In this section, we introduce a pulse wave quality assessment algorithm designed to eliminate data with poor pulse wave quality and provide guidance on the mobile phone's position. AF detection can only be reliably performed on signals that contain high-quality pulse waves, which can accurately reflect cardiac activity. If the signals are noisy or fail to capture high-quality pulse waves, the detection results will be unreliable. Therefore, it is crucial to filter out low-quality data to ensure effective daily AF detection and to guide users to position their smartphones where pulse waves can be detected more effectively. Building on the work in \cite{mmArrhythmia, waltz} and our analysis of real-world data, we propose two key metrics for assessing pulse wave quality: the Channel Stability Score $C$ and the Cardiac Band Spectrum Energy Ratio $\eta_{c}$.

\subsection{Channel Stability Score}
The signal quality is highly correlated with channel stability. To measure this, we calculate the similarities between the four channels. When the mobile phone is held stably and placed properly, the channels capture consistent pulse waves, leading to high similarity across the channels. Conversely, when severe motion artifacts or interference occur, the channels become unstable, and the pulse waves may fluctuate and contain a lot of noise. Given that noise has a random nature, the similarity in phase sequences between different channels will decrease. Thus, we use this similarity metric to assess channel stability.

The calculation of $C$ is formulated as follows: given four phase sequences, denoted as $\{\phi_{est}^{1}(t), \phi_{est}^{2}(t), \phi_{est}^{3}(t),$ $\phi_{est}^{4}(t)\}$, corresponding to the four different channels, the cosine similarity $S(m,n)$ between any two channels, $\phi_{est}^{m}(t)$ and $\phi_{est}^{n}(t)$, is computed as:
\begin{align}
S(m,n) = \frac{\sum_{t=1}^{T} \phi_{est}^{m}(t) \cdot \phi_{est}^{n}(t)}{ \sqrt{\sum_{t=1}^{T} \left( \phi_{est}^{m}(t) \right)^2} \cdot \sqrt{\sum_{t=1}^{T} \left( \phi_{est}^{n}(t) \right)^2}},
\end{align}
where $T$ represents the signal length over a 30-second time segment, and $m, n \in \{1, 2, 3, 4\}$. 

The \textbf{channel stability score}, denoted as \( C \), is defined as the average cosine similarity between all pairs of channels. Some channels may be affected by the frequency selection problem, leading them to capture weak or no pulse waves, resulting in low similarity with other channels. To prevent these channels from lowering the stability score, we exclude pairs with similarity below 60\% of the maximum similarity observed. This ensures that only stable channel pairs are considered in the final score, providing a more accurate reflection of overall channel stability. Formally, \( C \) is expressed as:
\begin{align}
C = \text{avg}\left( \left\{ S(m,n) \mid m \neq n \text{ and } S(m,n) \geq 0.6 \cdot \max_{m \neq n} S(m,n) \right\} \right),
\end{align}
where \( S(m,n) \) represents the cosine similarity between channel \( m \) and channel \( n \). The threshold \( 0.6 \cdot \max S(m,n) \) ensures that the score reflects only those pairs with a sufficiently high degree of correlation, providing a robust indicator of the overall channel stability.

\begin{figure}
  \centering
  \includegraphics[width=\linewidth]{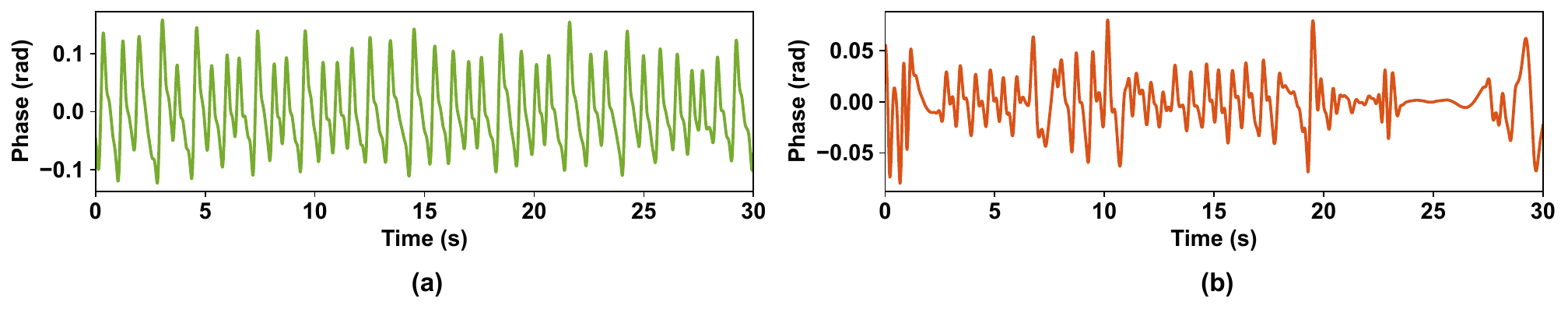}
  \caption{Examples of (a) high-quality pulse waves and (b) low-quality pulse waves.}
  \label{f: quality.example}
\end{figure}

\begin{figure}
  \centering
  \includegraphics[width=4.5in]{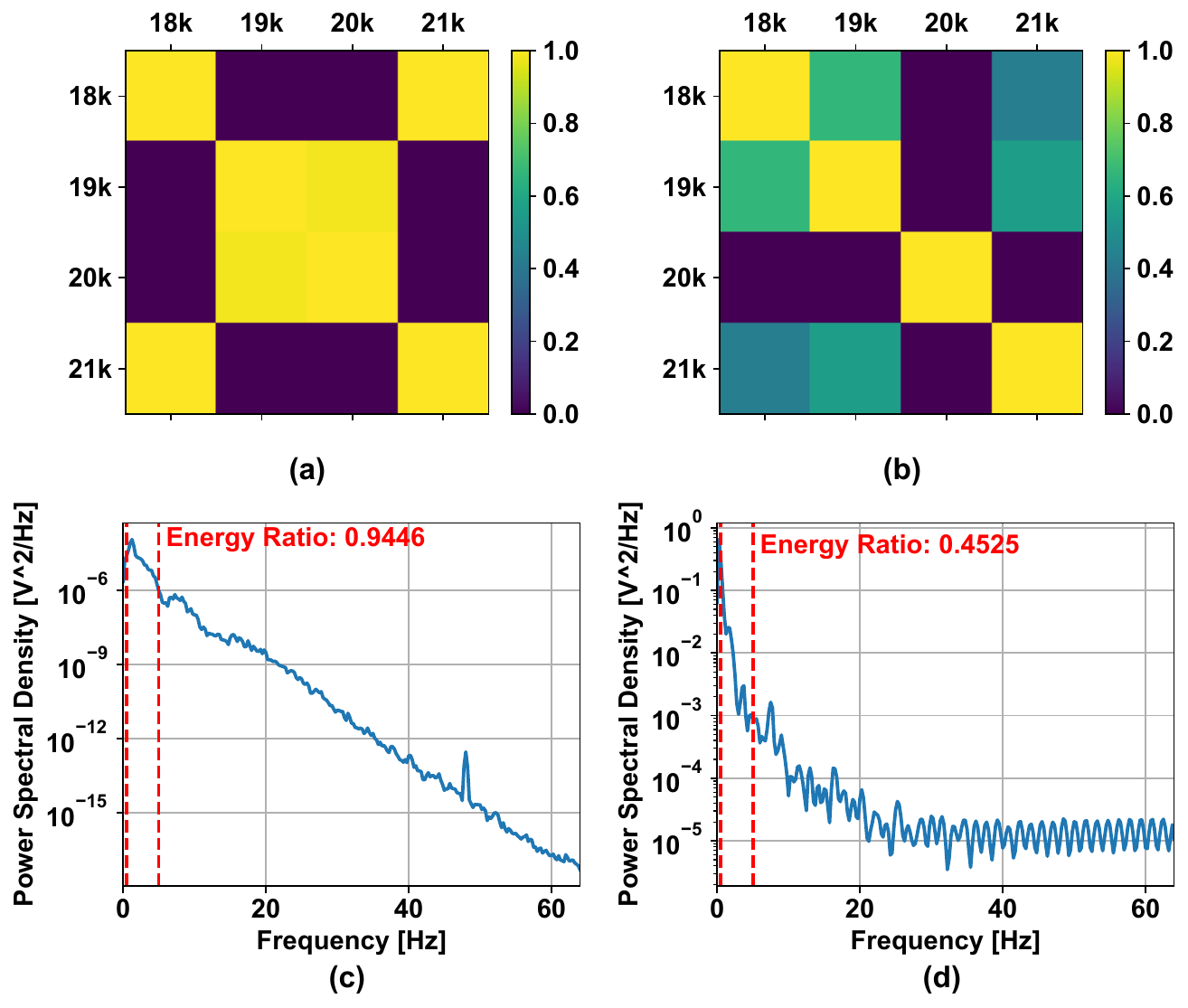}
  \caption{Difference of the pulse waves' similarity and power spectrum with different data qualities. \textbf{(a):} The cosine similarity matrix with high-quality pulse waves. \textbf{(b):} The cosine similarity matrix with low-quality pulse waves. \textbf{(c):} The power spectrum of high-quality data. \textbf{(d):} The power spectrum of low-quality data.}
  \label{f: quality.details}
\end{figure}

To exclude low quality data, we set 0.90 as the threshold of $C$ based on our experience, which means that any segment with a $C \leq 0.90$ is considered low-quality and removed from further analysis. Fig.~\ref{f: quality.example}(a) and Fig.~\ref{f: quality.example}(b) illustrate two examples of high- and low-quality signals, respectively. In Fig.~\ref{f: quality.details}(a) and Fig.~\ref{f: quality.details}(b), the cosine similarity matrices of the phase sequences in different carriers are depicted for both signals. Channels that are attenuated by frequency selectivity tend to have lower SNR, and therefore, certain channels exhibit low similarity with others. However, the overall similarity in Fig.~\ref{f: quality.details}(a) is significantly higher than that in Fig.~\ref{f: quality.details}(b), with $C$ of 0.99 compared to 0.54. The results indicate that $C$ is an effective metric for assessing data quality.

\subsection{Cardiac Band Spectrum Energy Ratio}
The normal resting heart rate typically ranges from 60 to 100 beats per minute \cite{resting_heartrate}, which corresponds to a frequency band of 0.5 Hz to 5 Hz. This range effectively captures the energy distribution of the pulse waves. However, brief, distorted periods within the 30-second segment may introduce errors in the analysis. To improve accuracy, the signal is divided into two halves, and the lower energy ratio between the two halves is used for further analysis.

The calculation of $\eta_{c}$ is given as follows: given four phase sequences for a 30-second signal, denoted as $\{\phi_{est}^{1}(t), \phi_{est}^{2}(t), $ $\phi_{est}^{3}(t), \phi_{est}^{4}(t)\}$, the \textbf{cardiac band spectrum energy ratio}, $\eta_{c}$, is defined as:
\begin{align}
    \eta_{c} = \max \left\{  \min \left(  \frac{e_{0 \sim 0.5T}^{i}}{E_{0 \sim 0.5T}^{i}}, \frac{e_{0.5T \sim T}^{i}}{E_{0.5T \sim T}^{i}}  \right) \mid i \in [1, 4]  \right\},
\end{align}
where $e_{0 \sim 0.5T}^{i}$ and $e_{0.5T \sim T}^{i}$ represent the spectrum energy of the pulse wave frequency band (0.5 to 5 Hz) within the first and second halves of the phase sequences $\phi_{est}^{i}(t)$. Similarly, $E_{0 \sim 0.5T}^{i}$ and $E_{0.5T \sim T}^{i}$ denote the total energy for the corresponding halves. 

To filter out low quality data, we set 0.70 as the threshold of $\eta_{c}$, meaning that any segment with $\eta_{c} \leq 0.70$ is considered low-quality. Fig.~\ref{f: quality.details}(c) and Fig.~\ref{f: quality.details}(d) present the power spectrum of the best channel from the two example signals. The band frequency ratio in the range of 0.5 Hz to 5 Hz in Fig.~\ref{f: quality.details}(c) is notably higher than in Fig.~\ref{f: quality.details}(d), with a ratio of 0.94 compared to 0.45. This difference is consistent with $\eta_{c}$, showing a value of 0.93 in Fig.~\ref{f: quality.details}(c) versus 0.37 in Fig.~\ref{f: quality.details}(d). The results demonstrate that $\eta_{c}$ is an effective metric which can reflect data quality.

%% file: Body/pulse_wave_purification.tex
\section{Pulse Wave Purification}
\label{s: pulse_wave_purification}
This section presents the key processing methods employed for pulse wave purification, which collectively enhance the quality of pulse wave signals.

\begin{figure}
  \centering
  \includegraphics[width=4.0in]{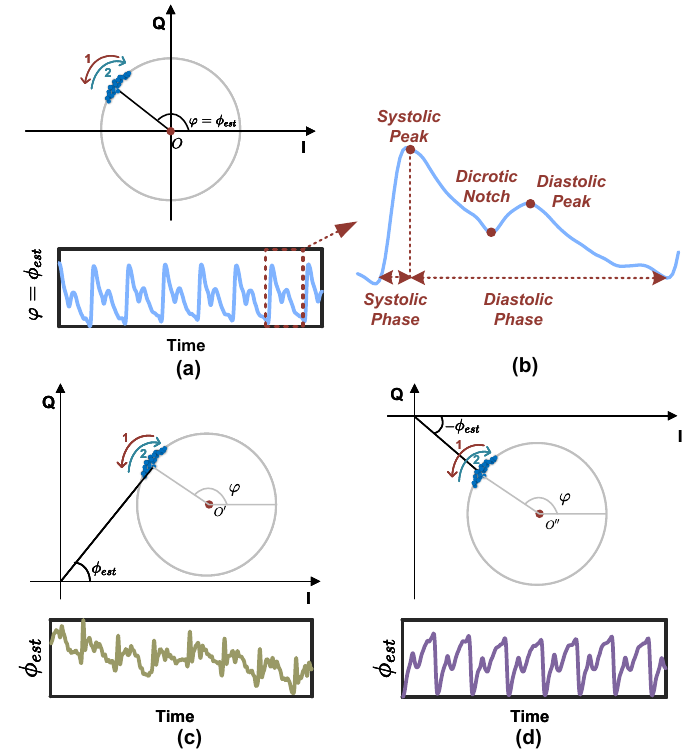}
  \caption{Illustration of the received pulse wave signal in the I/Q Domain. $\varphi$ and $\phi_{est}$ represent the cardiac-related phase and the calculated phase in practice, respectively. Arrow 1 and Arrow 2 represent the typical phase change direction in cardiac systolic period and diastolic period, respectively. \textbf{(a)}: The ideal case where the arc center is located at the coordinate origin, facilitating $\phi_{est}$ to be consistent with $\varphi$. \textbf{(b)}: Illustration of an example pulse wave. \textbf{(c)}: The worst case scenario where the arc center drifts, hindering the changes of $\varphi$ to be clearly shown in $\phi_{est}$ and leading to the distortion of the extracted pulse waves. \textbf{(d)}: A common case in which the drift arc center leads the $\phi_{est}$ to be in the inverted direction of $\varphi$.}
  \label{f: circle_rectification_intro}
\end{figure}

\subsection{Static Component Elimination}
\label{s: waveform_correction_and_amplification}
Due to the transmission through the device's internal structure, the received signals usually contain lots of static components, leading the pulse waveform to be distorted and noisy. To address this problem, we propose an efficient algorithm on the I/Q domain to eliminate static components.

Inspired by existing research works \cite{mmVib, mTrack, waltz}, the I and Q components of the received signal, discussed in Sec. \ref{s: pulse_wave_extraction}, form an arc in the I/Q domain. Ideally, the center of this arc is theoretically located at the origin of the coordinates. Consequently, the calculated phase changes correspond to actual channel state changes, reflecting pulse waves. As shown in Fig.~\ref{f: circle_rectification_intro}(a), $\varphi$ and $\phi_{est}$ represent the cardiac-induced phase and the calculated phase, respectively. The dots represent a series of sample points captured by the mobile phone, forming an arc in the I/Q domain. When the heart is in the systolic period, $\varphi$ typically increases, causing the sample points to move counterclockwise. Conversely, during the diastolic period, $\varphi$ decreases, leading the sample points to move clockwise. In this ideal scenario, the arc center is located at or very close to the origin, and $\phi_{est}$ accurately reflects the changes in $\varphi$.

However, in practice, the signals received by the microphone include not only the signals reflected from the wrist, but also the signals emitted directly from the speaker or reflected from other structures of the phone. This results in a static component in the received signal, leading to arc center drift~\cite{mTrack}. Arc center drift causes the received pulse waveform to be distorted and noisy. As shown in Fig.~\ref{f: circle_rectification_intro}(c), although the $\varphi$ changes are similar to Fig.~\ref{f: circle_rectification_intro}(a), and the movement pattern of the sampling points is similar, the drifting arc center causes the $\phi_{est}$ to be severely blurred. Meanwhile, arc center drift can lead to signal inversion, as shown in Fig.~\ref{f: circle_rectification_intro}(d). As a result, arc center drift prevents the $\phi_{est}$ from accurately reflecting changes in the channel state, thereby hindering the precise and reliable extraction of the pulse waves.

\begin{figure}
  \centering
  \includegraphics[width=5.5in]{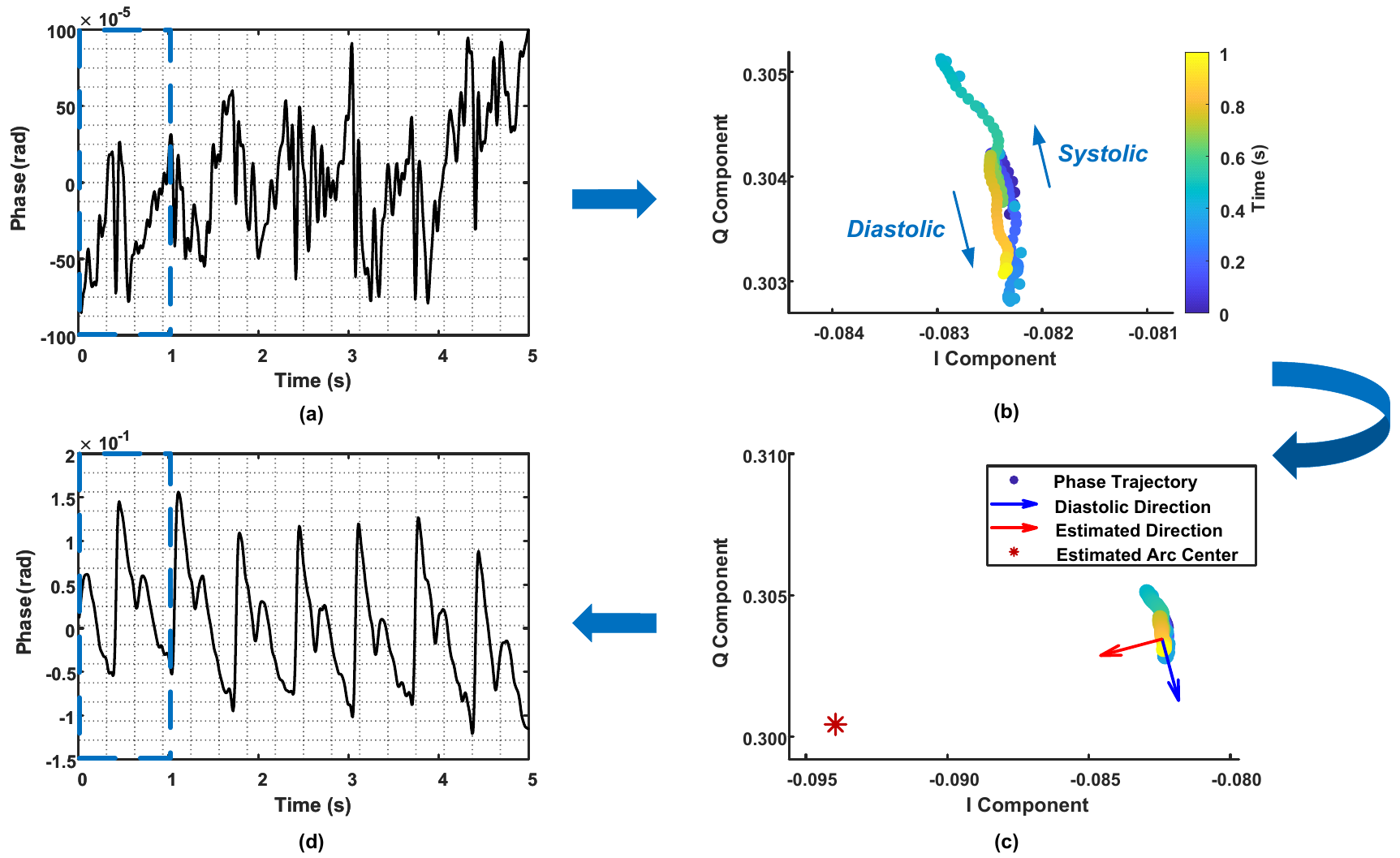}
   \caption{Illustration of Static Components Removal. \textbf{(a)}: Original pulse waves before Static Component Removal. \textbf{(b)}: Trajectory in the I/Q domain for the first second of (a). \textbf{(c)}: The trajectory movement direction in diastolic period (diastolic direction), estimated arc center direction, and estimated arc center of (b). \textbf{(d)}: Pulse waves after Static Component Removal.}
  \label{f: waveform_correction_and_amplification}
\end{figure}

Therefore, the challenge of \systemname is to eliminate the static components. Most previous studies have focused on finding the arc center from the geometric perspective, \ie, circular fitting \cite{waltz}. However, this approach is ineffective for \systemname. This is due to the fact that the vibration of the wrist induced by heartbeat is on the order of millimeters, which is too short compared to the wavelength of sound on the order of centimeters. This discrepancy tends to cause the fitted arc center to be outside the arc, resulting in an inverted phase. Since our system focuses on the changes of phase rather than its absolute values, we only need to obtain an estimated arc center close to the geometric arc center to amplify the phase changes. To achieve this, we identify two key observations to help estimate the arc center.

\textbf{Observation 1: Linear Approximation of the I/Q Arc}. The arc in the I/Q domain is sufficiently short to be approximated as a straight-line segment. Given the minute impact of the pulse wave on the channel, the phase remains nearly constant along the arc. Consequently, the trajectory of the sampling point can be approximated as a short straight-line segment.

\textbf{Observation 2: Directional Velocity Difference during Cardiac Cycle}. The trajectory generally moves clockwise along the arc with slower velocity during the diastolic period and in the opposite direction with faster velocity during the systolic period. Fig.~\ref{f: waveform_correction_and_amplification}(b) illustrates the trajectory in the I/Q domain for the first second of Fig.~\ref{f: waveform_correction_and_amplification}(a), where the trajectory points nearly form a line. At the same sampling rate, the distances between the sample points moving from the bottom-right to the top-left are generally shorter and more densely packed than those moving in the reverse direction, indicating slower velocity. This observation suggests that moving from the bottom right to the upper left is systolic and vice versa for the diastolic phase of the heart. Since the points always rotate along the clockwise direction around the arc center in the diastolic period, we can estimate the direction of the center of the arc by calculating the direction of movement of the sample points in diastole.



Based on the above observations, we propose an \textit{Arc Center Estimation Algorithm} to eliminate the impact of static components, which consists of the following three steps:

\textbf{Step 1: Trajectory Moving Primary Direction Determination.} Given a sequence of sampling points \( \vec{P}_1, \vec{P}_2, \ldots, \vec{P}_n \) in the I/Q domain using the algorithm introduced in Sec.~\ref{s: pulse_wave_extraction}, we first compute the trajectory vectors \( \vec{v}_i \) as the displacement between consecutive points:
\begin{eqnarray}
\vec{v}_i = \vec{P}_{i+1} - \vec{P}_i \quad \text{for} \quad i = 1, 2, \ldots, n-1.
\end{eqnarray}
Then, we obtain the direction of the line segment by employing PCA on trajectory vectors: 
\begin{eqnarray}
\vec{pc}_1, \vec{pc}_2 \longleftarrow PCA(\vec{v}_1, \vec{v}_2, \ldots, \vec{v}_{n-1}),
\end{eqnarray}
where $\vec{pc}_1$ and $\vec{pc}_2$ represent the first principal component and the second principal component, respectively. The first principal component $\vec{pc}_1$ represents the direction of the line segment, as it identifies the direction with the greatest variance among the trajectory vectors, which indicates the primary direction of movement.

\textbf{Step 2: Diastolic Trajectory Moving Direction Determination.} We project each trajectory vector \( \vec{v}_i \) onto the first principal component \( \vec{pc}_1 \) to obtain the projected vectors:
\begin{eqnarray}
\vec{p}_i = \frac{\vec{v}_i \cdot \vec{pc}_1}{\|\vec{pc}_1\|^2} \vec{pc}_1.
\end{eqnarray}
To determine the direction that the sample points move along in the diastolic period, we calculate the average magnitude of the projected vectors \( \vec{p}_i \) for each side of the line segment. Define \( M_+ \) and \( M_- \) as the sets of magnitudes of projected vectors in the positive and negative directions of \( \vec{pc}_1 \), respectively:
\begin{eqnarray}
M_+ = \{ \|\vec{p}_i\| \mid \vec{v}_i \cdot \vec{pc}_1 > 0 \}, \\
M_- = \{ \|\vec{p}_i\| \mid \vec{v}_i \cdot \vec{pc}_1 < 0 \}.
\end{eqnarray}
We then compute the average magnitudes \( \bar{M}_+ \) and \( \bar{M}_- \):
\begin{eqnarray}
\bar{M}_+ = \frac{1}{|M_+|} \sum_{\vec{p}_i \in M_+} \|\vec{p}_i\|, \\
\bar{M}_- = \frac{1}{|M_-|} \sum_{\vec{p}_i \in M_-} \|\vec{p}_i\|.
\end{eqnarray}
According to Observation 2, the direction in which sampling points move with slower velocity represents the direction of the movement of the sample points in cardiac diastole. The direction of slower movement is given by:
\begin{eqnarray}
\vec{d} = 
\begin{cases} 
\vec{pc}_1 & \text{if } \bar{M}_+ < \bar{M}_-, \\
-\vec{pc}_1 & \text{if } \bar{M}_+ > \bar{M}_-.
\end{cases}
\end{eqnarray}

\textbf{Step 3: Arc Center Estimation.} We estimate the arc center direction by rotating the determined direction \( \vec{d} \) clockwise by 90 degrees, ensuring that the sample points move along the clockwise direction of the estimated arc center during the diastolic period:
\begin{eqnarray}
\vec{d}_{\text{arc}} = \begin{pmatrix}
0 & 1 \\
-1 & 0
\end{pmatrix} \vec{d}.
\end{eqnarray}
Conventionally, we can calculate the radius by circular fitting or finding the intersection of the perpendicular bisectors of two chords from three points on the arc. However, it is either time-consuming or not suitable for extremely short arc. Again, we do not pursuit the accurate arc center in geometry, but an estimated arc center that can amplify the pulse wave. Therefore, we calculate the distance between the two furthest sampling points \( \vec{P}_i \) and \( \vec{P}_j \) to estimate the radius. And the position of the estimated circle center \( \vec{C} \) is then given by:
\begin{eqnarray}
d_{\text{max}} = \max_{i, j} \|\vec{P}_i - \vec{P}_j\|, 
\end{eqnarray}
\begin{eqnarray}
\vec{C} = \frac{1}{n} \sum_{i=1}^n \vec{P}_i + \eta \times d_{\text{max}} \cdot \vec{d}_{\text{arc}},
\end{eqnarray}
where $\eta$ is an empirical coefficient to adjust the radius of the arc, which is set to be 5 in \systemname. Once the arc center is estimated, we can use the center to rectify the sample points $\hat{\vec{P}}_{i}$, eliminating the influence of the drift of the arc center:
\begin{eqnarray}
    \hat{\vec{P}}_{i} = \vec{P}_{i} - \vec{C}.
\end{eqnarray}

Due to the dynamic nature of the arc center, we continuously adjust the arc center rather than making a single correction. We exploit a running window with 0.5 s as the step size and 2.5 s as the window size. This method, while effective, still introduces some discontinuities in the results. To mitigate these discontinuities, we only apply rectification when the angle difference between the estimated and actual arc center directions exceeds a threshold of $\frac{\pi}{6}$. This approach helps maintain continuity unless a significant correction is required. In order to further reduce the discontinuity, the junction is smoothed if rectification is performed. Fig.~\ref{f: waveform_correction_and_amplification}(c) details the process of estimating the arc center for the initial two-second trajectory shown in Fig.~\ref{f: waveform_correction_and_amplification}(b). Finally, Fig.~\ref{f: waveform_correction_and_amplification}(d) displays the outcome after applying our \textit{Static Component Elimination} to Fig.~\ref{f: waveform_correction_and_amplification}(a), resulting in a non-reversed, smoother, and significantly clearer waveform.

\subsection{Frequency Selection}
\label{s: pulse_wave_purification.frequency_selection}
To minimize computational costs in subsequent analyses, we aim to identify and select the frequency that produces the highest quality pulse wave signal. \systemname assesses the quality of pulse waves at each frequency using two primary metrics:
\begin{itemize}
    \item \textbf{Average PCA Explained Variance Ratio $P$.} This metric is used to quantify the proportion of variance captured by the first principal component during PCA analysis for each window of the signal. For each window $i$, the explained variance ratio of the first principal component is denoted as $\lambda_1^{(i)}$. The average PCA explained variance ratio across all $N$ windows is then calculated as $ P = \frac{1}{N} \sum_{i=1}^{N} \lambda_1^{(i)}$, where $P$ represents the average explained variance ratio, $N$ is the total number of windows, and $\lambda_1^{(i)}$ is the explained variance ratio of the first principal component for the $i$-th window. A higher value of $P$ indicates less noise in the signal, as discussed in Section~\ref{s: waveform_correction_and_amplification}.

    \item \textbf{Band Spectrum Energy Ratio $\eta_{b}$}. This metric is defined as $\eta_{b} = \frac{e_b}{e_{total}}$, where $e_b$ is the spectrum energy of pulse wave frequency band $0.5 \sim 5$ Hz, and $e_{total}$ is the total energy. Notably, the spectrum energy of band $0 \sim 0.2$ Hz is ignored when calculating $e_{total}$ because of constant components in Fast Fourier Transform while calculating the spectrum. 
\end{itemize}
 After computing and averaging the scores from these two metrics across all frequencies, \ie, $S=(P+\eta_{b})/2$, we select the frequency with the highest $S$. The pulse wave signals corresponding to this frequency are then chosen for further processing.

\begin{figure}[t]
  \centering
  \includegraphics[width=4.5in]{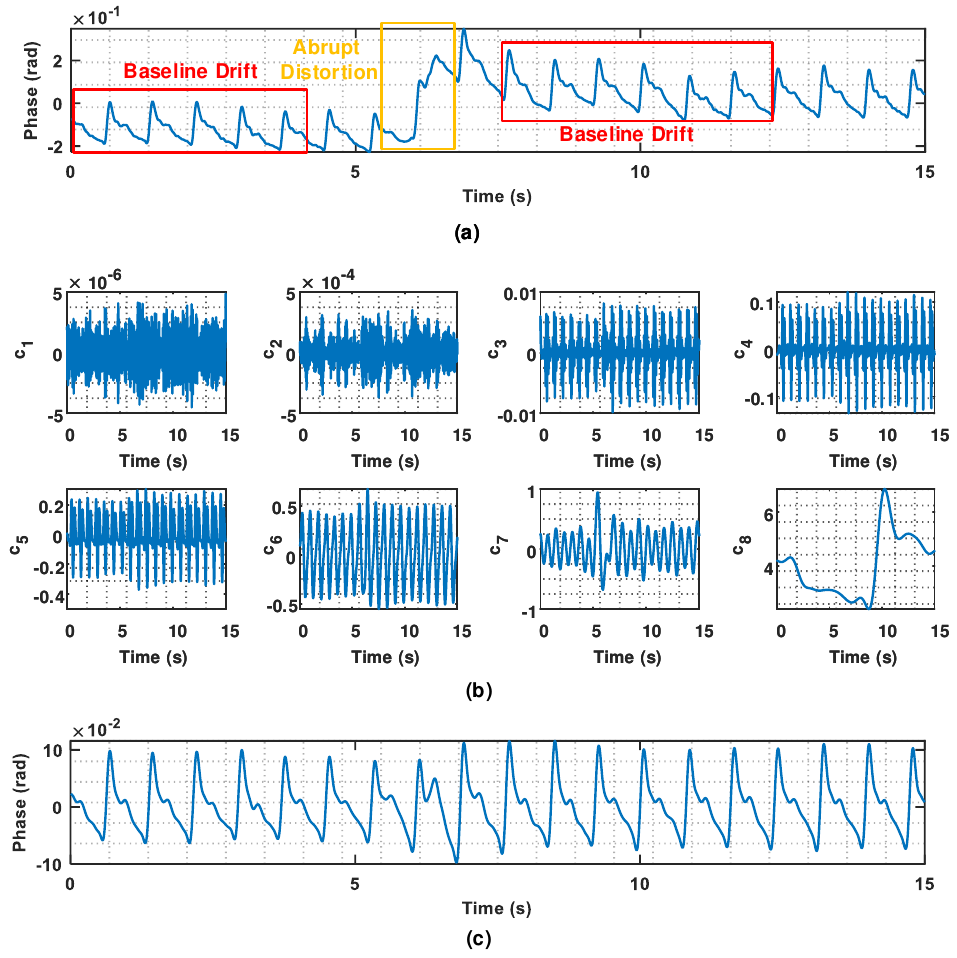}
  \caption{Illustration of Motion Artifacts Removal. \textbf{(a)}: Original pulse waves before Motion Artifacts Removal. \textbf{(b)}: SWT Composition Result of (a). \textbf{(c)}: Pulse waves after Motion Artifacts Removal.}
  \label{f: interference_removal}
\end{figure}

\subsection{Motion Artifacts Removal}
\label{s: pulse_wave_purification.interference_removal}
When using a mobile phone for atrial fibrillation detection, the phone's speaker and microphone are located near the radial artery skin at the wrist. However, there may be some slight relative motion between the hand and the device because the device is handheld. Those motion artifacts can introduce baseline drift and abrupt distortion into the received signal, which complicates or disturbs the detection of AF. Therefore, the removal of motion artifacts is essential for purifying pulse waves. 

Inspired by the work of Wang \etal \cite{swt1}, \systemname performs SWT on extracted pulse waves to remove motion artifacts. In particular, \systemname uses Coif5 as the wavelet as it has greater advantages in time-frequency localization and smoothing. Since valid pulse waves typically fall within the frequency range of 0.5 to 5 Hz, a 7th-order decomposition is performed on the extracted pulse waves. This decomposition yields eight coefficient sequences $\left\{c_1, c_2, \ldots, c_8\right\}$ corresponding to different frequency ranges: 32.0 to 64.0 Hz, 16.0 to 32.0 Hz, 8.0 to 16.0 Hz, 4.0 to 8.0 Hz, 2.0 to 4.0 Hz, 1.0 to 2.0 Hz, 0.5 to 1.0 Hz, and 0 to 0.5 Hz, progressively halving the frequency range in each step. Subsequently, the pulse waves are reconstructed using inverse stationary wavelet transformation with wavelet coefficients $c_4$ to $c_7$, corresponding to the frequency range from 8.0 Hz to 0.5 Hz. 

Fig.~\ref{f: interference_removal}(b) illustrates the composition result of the pulse waves depicted in Fig.~\ref{f: interference_removal}(a), indicating that baseline drift information mainly resides in $c_7$ and $c_8$, pulse wave information is primarily decomposed into $c_3$ to $c_6$, and high-frequency noise is predominantly found in $c_1$ to $c_2$. The reconstructed pulse waves, depicted in Fig.~\ref{f: interference_removal}(c), appears more stable and smoother compared to the raw pulse waves, thereby facilitating accurate AF detection, as discussed in the subsequent section.

%% file: Body/AF_detection.tex
\begin{figure}
  \centering
  \includegraphics[width=5in]{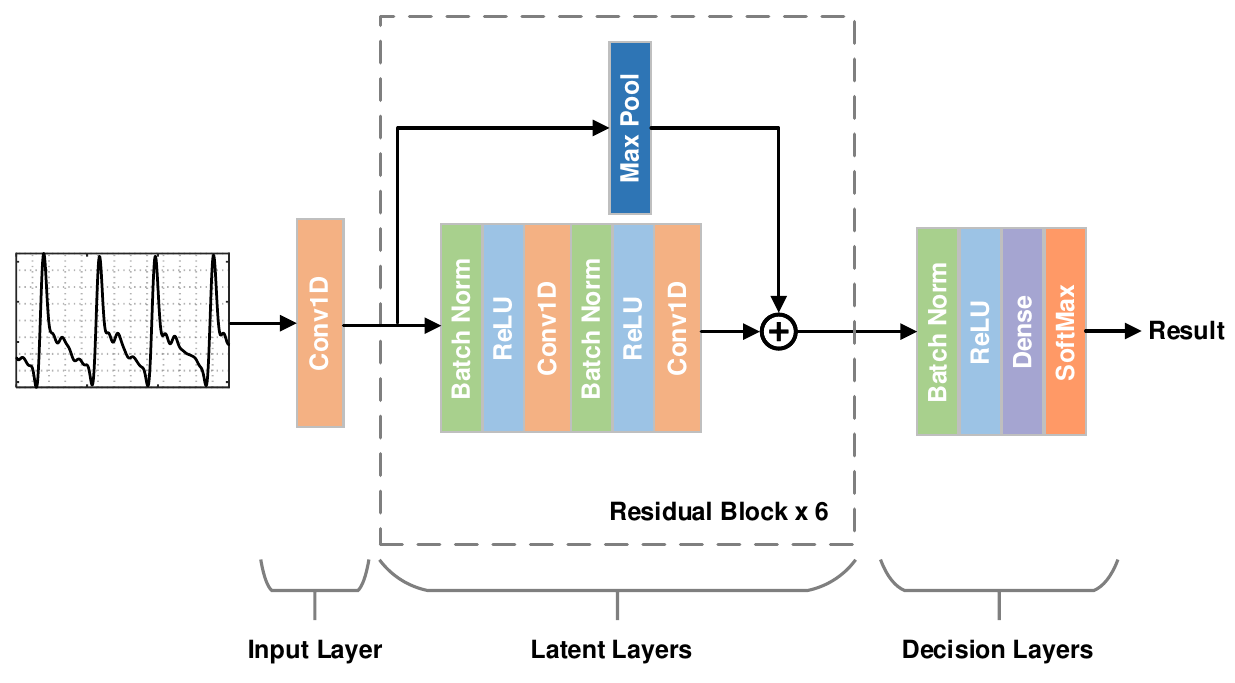}
  \caption{The architecture of AF detector.}
  \vspace{-0.1in}
  \label{f: af_detection_model_overview}
\end{figure}

\begin{table}
    \centering
    \caption{Parameters of AF detector.}
    \begin{tabular}{l|rrrrr}
        \toprule
        Layer             & Kernel Size & \multicolumn{1}{r}{Stride} & Output Channel \\
        \midrule
        Input Conv1D     & 32          & 1                           & 16             \\ 
        Residual Block 1 & 32          & 1                           & 16             \\
        Residual Block 2 & 32          & 4                           & 16             \\ 
        Residual Block 3 & 32          & 1                           & 16             \\
        Residual Block 4 & 32          & 4                           & 16             \\ 
        Residual Block 5 & 32          & 1                           & 16             \\
        Residual Block 6 & 32          & 4                           & 16             \\ 
        \bottomrule
    \end{tabular}
  \label{tab: af_detection_model_parameters}
\end{table}

\section{AF Detector}
\label{s: af_detection}
To implement an efficient and accurate AF detector, we propose a ResNet-based classification network \cite{he2015deepresiduallearningimage}. Fig.~\ref{f: af_detection_model_overview} illustrates the architecture of \systemname AF detector, designed to be lightweight and efficient for mobile devices. The AF detector processes purified one-dimensional (1D) pulse wave segments as input, which undergo Z-score normalization before being fed into the detector. The input layer contains a single 1D convolutional block (Conv1D), expanding the initial single channel into 16 channels and extracting preliminary features. The latent layers consist of 6 residual blocks that work together to uncover the hidden information in the pulse wave signals. Starting from the second residual block, the detector performs a four-fold downsampling every two residual blocks, which effectively reduces the computational cost and increes the receptive field. The decision layers comprise a batch normalization layer, a rectified linear unit~(ReLU) activation layer, a dense layer, and a softmax layer, which fuse the features extracted by the latent layers and make the final decision. Detailed parameters are listed in Table~\ref{tab: af_detection_model_parameters}.

%% file: Body/implementation.tex
\begin{figure}
  \centering
  \includegraphics[width=5.2in]{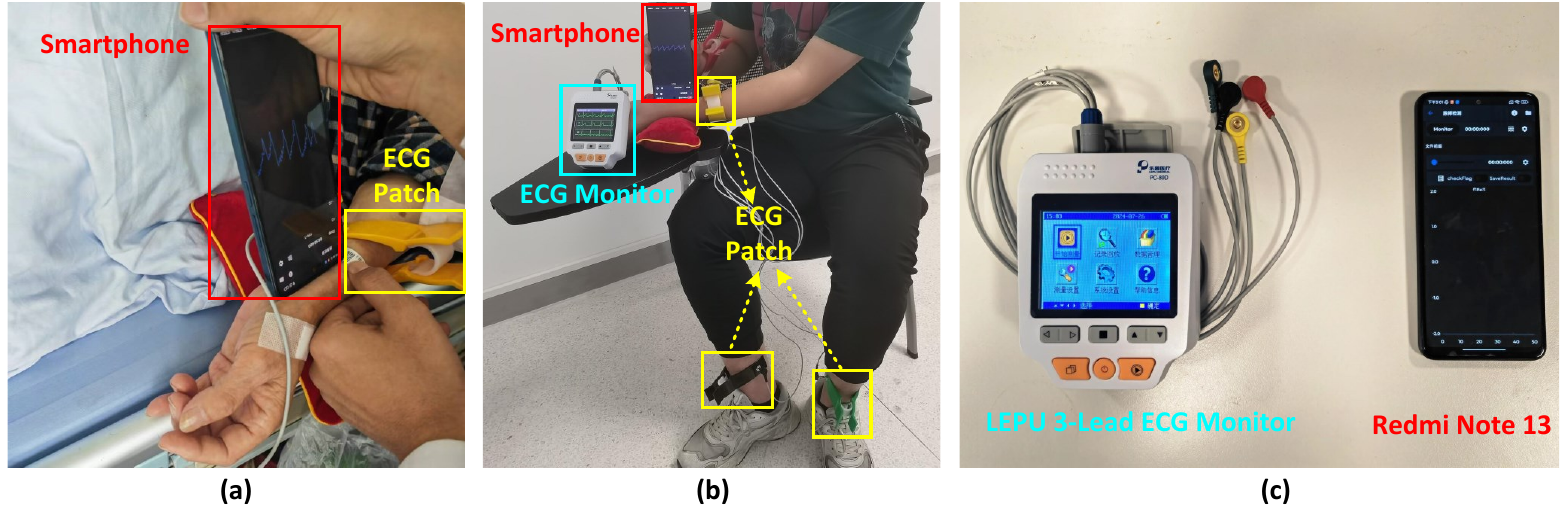}
  \caption{\textbf{(a)} Data collection in clinical scenario. \textbf{(b)} Data collection in daily scenario. \textbf{(c)} LEPU 3-Lead ECG Monitor and Redmi Note 13.}
  \vspace{-0.2in}
  \label{f: setup}
\end{figure}

\section{Implementation}
\label{s: implementation}
Our system is implemented on a Redmi Note 11 Pro smartphone, utilizing its built-in top speakers and microphones, as shown in Fig.~\ref{f: setup}(c). Acoustic data are collected using the smartphone and subsequently transferred to a desktop computer with 16 GB memory and an AMD Ryzen 7 3700X 8-core processor for further processing. The smartphone records at a sampling rate of 48,000 Hz. Reference ECG data are obtained from the LEPU Three-lead ECG Monitor PC-80D, with a sample rate of 150 Hz. Both acoustic and ECG data are annotated using Label Studio, with ECG serving as the ground truth reference. Algorithms detailed in Sec. \ref{s: pulse_wave_probing}, Sec. \ref{s: pulse_wave_extraction}, and Sec. \ref{s: pulse_wave_purification} are implemented with MATLAB 2023b, and Algorithms detailed in Sec. \ref{s: pulse_wave_quality_screening} are implemented with Python 3.10. The ECG data and acoustic data are aligned by their recording start timestamps, with an error margin of less than 10 seconds. For Sec. \ref{s: af_detection}, we implement our model with PyTorch \cite{pytorch-library} on the basis of the work by Hong et al. \cite{hong_resnet1d} and train it on a server with NVIDIA GeForce GTX 3090 GPU and 64 GB memory. We adopt Adam as the optimizer with an initial learning rate of 0.001. The training proceeds for 50 epochs with possible early stopping.

%% file: Body/evaluation.tex
\section{Evaluation}
\label{s: evaluation}
\subsection{Experiment Setup}
\subsubsection{Data Collection}
We recruit 23 participants (18 males, 5 females) aged between 20 and 89 years, with an average age of 45.5 years, from our university and its affiliated hospital. Among the participants, 9 are diagnosed with AF, while 13 are non-AF participants. Unless otherwise specified, all experiments are conducted in a quiet conference room or ward, as shown in Fig.~\ref{f: setup}(a) and Fig.~\ref{f: setup}(b). For the performance evaluation of \systemname, we collect acoustic data using the top microphone and speakers of the Redmi Note 11 Pro smartphone, with a sampling rate of 48,000 Hz. Additionally, we collect ECG data using the  LEPU three-lead ECG Monitor PC-80D, with a sampling rate of 150 Hz. Each piece of data, whether acoustic or ECG, is recorded for a duration of 30 seconds. Participants are asked to engage in the experiments for approximately 30 minutes each. We have collected 1104 pieces of valid data in total. The entire study is conducted under the supervision of our institution's Institutional Review Board \footnote{SUSTech IRB No. 20240084.}.

\subsubsection{Ground Truth}
As mentioned earlier, the gold standard for diagnosing atrial fibrillation is ECG. Therefore, we utilize ECG as our ground truth for the collected data. While acquiring signals at the radial artery, we simultaneously use a commercially available portable ECG monitor from LEPU three-lead ECG Monitor PC-80D to perform ECG detection. This device has obtained certification from the China National Medical Products Administration. We align the data between the pair of smartphones and the ECG patch using timestamps. The acquired 30-second recordings are then reviewed and labeled as either AF or non-AF by expert physicians.

\subsubsection{Evaluation Metrics}
We employ a confusion matrix to provide the overall performance of our system. We define AF records as positive samples while non-AF records as negative samples. Based on the elements in the confusion matrix, \ie, True Negatives (TN), True Positives (TP), False Negatives (FN), and False Positives (FP), the following comprehensive metrics are utilized:

\begin{itemize}
    \item \textit{Accuracy}: It measures the overall correctness of the classification by calculating the ratio of correctly classified samples to the total number of samples. It can be formulated as $\frac{TN + TP}{TN + TP + FN + FP}$.
    \item \textit{Precision}: It quantifies the proportion of correctly predicted positive samples out of all predicted positive samples. Precision indicates the system's ability to minimize false alarms to non-AF users. It can be formulated as $\frac{TP}{TP + FP}$.
    \item \textit{Recall (Sensitivity)}: It assesses the proportion of correctly predicted positive samples out of all actual positive samples. Recall indicates the system's ability to alarm genuine AF patients. It can be formulated as $\frac{TP}{TP + FN}$.
    \item \textit{Specificity}: It indicates the fraction of true negatives to all real negative samples, \ie, $\frac{TN}{TN+FP}$. High specificity means that a normal heartbeat will not be misdiagnosed as AF.
    \item \textit{$F_{1}$ Score}: It is calculated as $F_{1} = 2* \frac{precision*recall}{precision+recall}$. It considers both precision and recall with the range of~[0,~1].
\end{itemize}

Although accuracy is a common metric for evaluating performance, we do not use it here due to issues with dataset imbalance. Instead, we use the F1 score as our primary performance metric. During training, we select the model from the epoch that achieves the highest F1 score in the validation phase as the final model.

\subsection{Overall Performance}
\label{s: evaluation.overall_performance}

\begin{figure}[htbp]
  \centering
  \begin{minipage}{0.48\textwidth}
    \centering
    \includegraphics[width=2in]{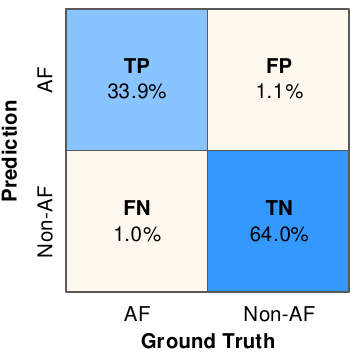}
    \caption{Averaged confusion matrix of 6-fold test results for overall performance evaluation.}
    \label{fig: confusion_matrix}
  \end{minipage}
  \hfill
  \begin{minipage}{0.48\textwidth}
    \centering
    \includegraphics[width=2.3in]{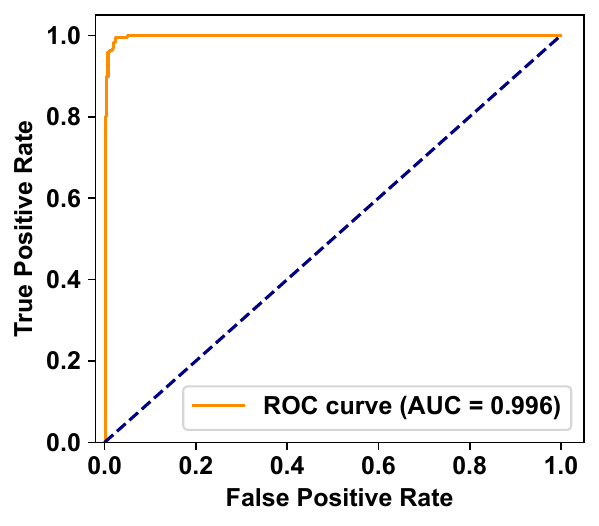}
    \caption{ROC curve for overall performance evaluation.}
    \label{fig: roc_curve}
  \end{minipage}
\end{figure}

For the overall evaluation, we adopt 6-fold cross-validation to minimize the impact of dataset distribution. This procedure involves partitioning the data into six equal parts, where during each validation step, a different fold is used as the test set, another randomly selected fold serves as the validation dataset, and the remaining parts constitute the training set. We record and accumulate the counts of TP, TN, FP, and FN from the test phase of each fold. These values are then utilized to calculate the average evaluation metrics across all folds. The dataset for overall evaluation comprises data from 19 participants (4 females and 15 males), all recorded in quiet environments.

Fig.~\ref{fig: confusion_matrix} illustrates the confusion matrix, comprising the consolidated results from all subjects involved. Our system achieves an average accuracy of 97.9\%, recall of 97.2\%, precision of 96.8\%, specificity of 98.3\% and an F1 score of 97.0\%. Fig.~\ref{fig: roc_curve} displays the receiver operating characteristic (ROC) curve, which shows the balance between the true positive rate (recall) and the false positive rate, with \systemname achieving an area under the curve (AUC) of 0.996. These results indicate that \systemname can effectively distinguish between patients with AF and those without.

\subsection{User-independent Performance}

\label{s: evaluation.user_independence}
\begin{figure}
  \centering
    \includegraphics[width=5.5in]{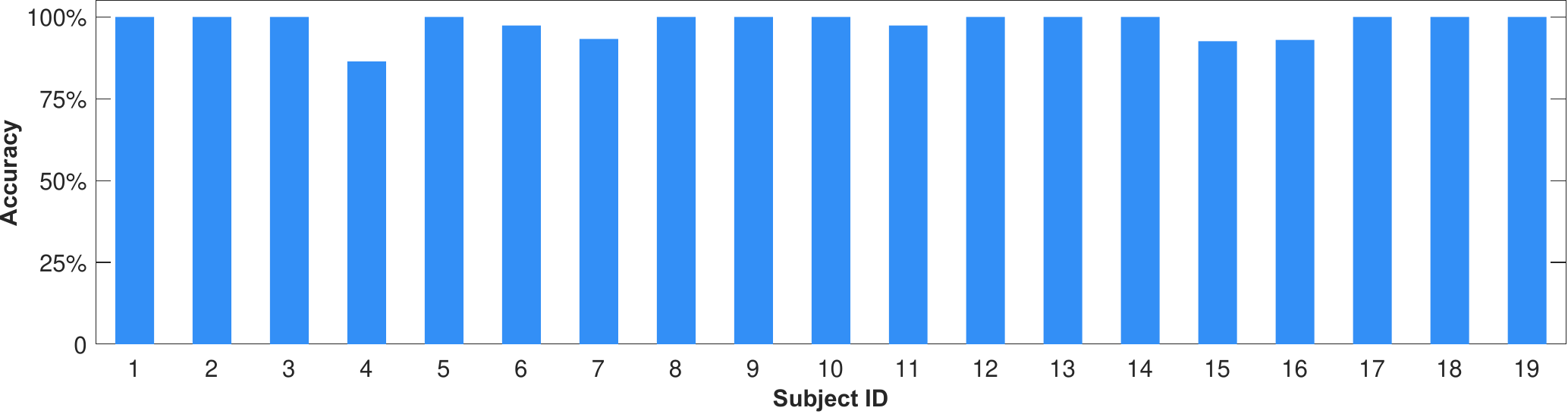}
  \caption{Result of leave-one-out validation of AF detection.}
  \label{f: user_independence}
\end{figure}

To evaluate the generalization capability of \systemname, we implement leave-one-out validation in a user-independent experimental setting. In this approach, one participant is held out as the test set, while the remaining subjects are randomly divided into an 80\% training set and a 20\% validation set. Fig.~\ref{f: user_independence} presents the accuracy bar plot for 20 subjects, where each subject can be considered an unseen participant. The average accuracy across the 19 subjects is 97.9\% with a standard deviation of 3.82\%. Notably, 15 out of 19 subjects exhibit high accuracy rates above 95\%, while only subject 4 records an accuracy below 95\%. The poor performance of subject 4 may be attributed to the limited amount of data available, which may not have been sufficient to mitigate the impact of system errors. Overall, \systemname demonstrates strong generalizability, effectively detecting AF in most unseen subjects.

\subsection{Effectiveness of Pulse Wave Quality Assessment} 
\begin{figure}
  \centering
  \includegraphics[width=4.5in]{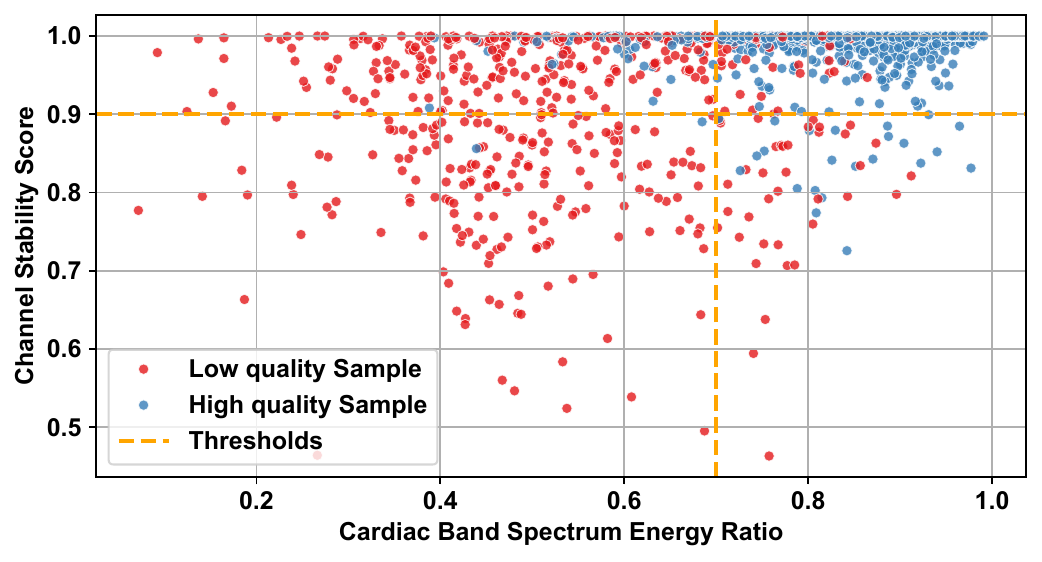}
  \caption{The scatter plot of cardiac band spectrum energy ratio and channel stability score for evaluation of \textit{Pulse Wave Quality Assessment}.}
  \label{f: evaluation.cardiac_filter}
\end{figure}
To effectively filter out low-quality data for daily AF detection, we introduce \textit{Pulse Wave Quality Assessment} as detailed in Sec.~\ref{s: pulse_wave_quality_screening}. We assess the module's performance by computing the metrics in the aforementioned section and calculating the ratio of low-quality data that are successfully filtered out and high-quality data that are preserved. The data quality is meticulously labeled by two technical experts specializing in cardiac data analysis, under the supervision of cardiologists to ensure clinical relevance and accuracy. The result shown in Fig.~\ref{f: evaluation.cardiac_filter} illustrates that most high-quality data exhibit a channel stability score above 0.9 and a cardiac band spectrum energy ratio above 0.7, clustering in the top right corner. With established thresholds of 0.9 for stability score and 0.7 for energy ratio, our system successfully filters out 91.7\% of noise segments while preserving 87.9\% of valid segments, which demonstrates the balanced performance of the pulse wave quality filter. The developed filter is also designed to provide real-time guidance during data collection, offering instant prompts to the user. However, during the collection of our dataset, this real-time application was not implemented, resulting in a higher proportion of invalid data compared to what would be expected in actual usage scenarios.

\subsection{Ablation Study}
\label{s: evaluation.ablation_study}

\begin{figure}
  \centering
    \includegraphics[width=4.5in]{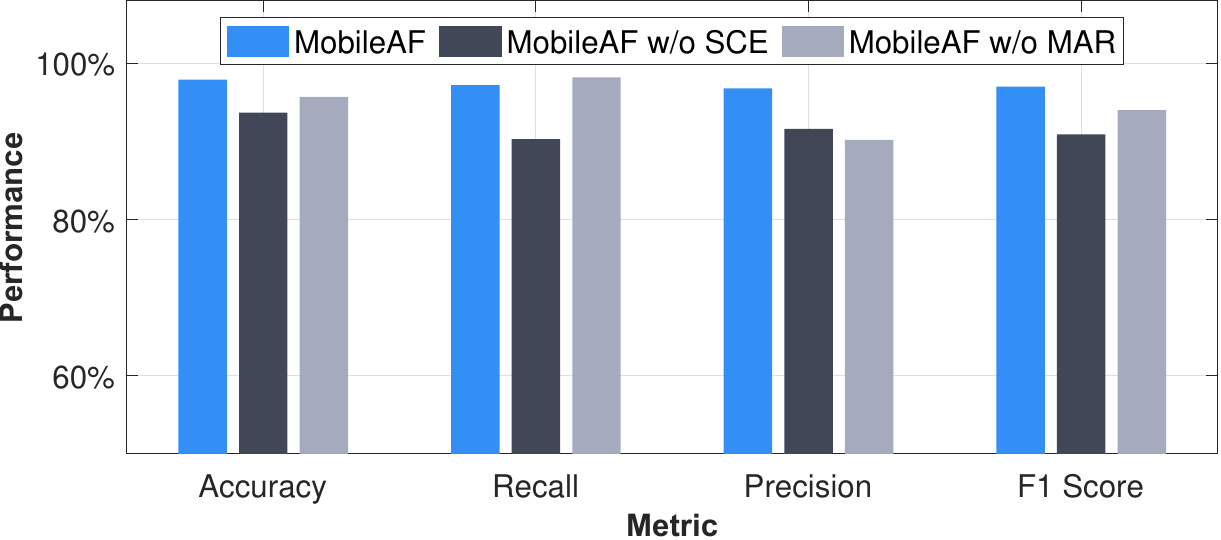}
\caption{System performance comparison of removing \textit{Static Components Elimination (SCE)} module or \textit{Motion Artifacts Removal (MAR)} module. }
\label{f: abalation_WCA_SWT}
\end{figure}

We conduct an ablation study to evaluate the effectiveness of the two key components of our system. To address the ambiguity and inversion caused by static components, we introduce \textit{Static Components Elimination}, as detailed in Sec.~\ref{s: waveform_correction_and_amplification}. Additionally, to mitigate the impact motion artifacts, we employ \textit{Motion Artifacts Removal}, as described in Sec.~\ref{s: pulse_wave_purification.interference_removal}. To assess the efficacy of these two approaches, we conduct a 6-fold cross-validation on the dataset discussed in Sec.~\ref{s: evaluation.overall_performance}. Specifically, we evaluate the dataset without applying \textit{Static Components Elimination} or \textit{Motion Artifacts Removal}.

The comparative results are illustrated in Fig.~\ref{f: abalation_WCA_SWT}. When trained and evaluated without \textit{Static Components Elimination (SCE)}, we observe a significant decrease in accuracy from 97.9\% to 93.7\%, a decline in precision from 97.2\% to 90.3\%, a reduction in recall from 96.8\% to 91.6\%, and a drop in the F1 score from 97.0\% to 90.9\%. Similarly, when trained and evaluated without \textit{Motion Artifacts Removal (MAR)}, there is a decrease in accuracy from 97.9\% to 95.7\%, a decline in precision from 96.8\% to 90.2\%, and a reduction in the F1 score from 97.0\% to 94.0\%. The recall is slightly higher while the other three metrics are lower, indicating that the AF detector may mistakenly classify some data with motion artifacts as AF data. These findings underscore the critical role of \textit{Static Components Elimination} in substantially enhancing the performance of \systemname.

\begin{table}[]
\caption{Performance of AF detectors implemented with different network structures.}
\label{tab: model_effectiveness}
\begin{tabular}{@{}llllll@{}}
\toprule
\multirow{2}{*}{Reference Work}    & \multirow{2}{*}{Network Structure} & \multicolumn{4}{l}{Test Performance}                              \\ \cmidrule(l){3-6} 
                                   &                                  & Accuracy       & Recall         & Precision      & F1 Score       \\ \midrule
Nurmaini et al. \cite{model_cnn}   & CNN                              & 88.9\%          & 91.7\%          & 79.5\%          & 85.2\%          \\
Zhang et al. \cite{model_lstm_cnn} & LSTM + CNN                       & 92.3\%          & 82.4\%          & 94.7\%          & 88.1\%          \\
Jin et al. \cite{model_ac_lstm}    & CNN + LSTM + Attention           & 93.1\%          & 94.4\%          & 94.7\%          & 90.5\%          \\
\textit{\textbf{Ours}}             & \textbf{ResNet}                  & \textbf{97.9\%} & \textbf{97.2\%} & \textbf{96.8\%} & \textbf{97.0\%} \\ \bottomrule
\end{tabular}
\end{table}

\subsection{Effectiveness of ResNet-based AF Detector}

As discussed in Sec.~\ref{s: af_detection}, we design a ResNet-based AF detector to achieve accurate and efficient AF detection. To evaluate its effectiveness, we implement three additional AF detectors using different network structures based on state-of-the-art works and test them with 6-fold cross-validation, as detailed in Sec.~\ref{s: evaluation.overall_performance}. The results, shown in Table~\ref{tab: model_effectiveness}, highlight the comparative performance.

The first AF detector, based on the work of Nurmaini et al. \cite{model_cnn}, uses a convolutional neural network (CNN). The original model has 13 CNN layers, which is too deep for \systemname, resulting in an F1 score below 70\%. We reduce the complexity by removing the last three layers, leaving the 10 layers. The second AF detector, derived from Zhang et al. \cite{model_lstm_cnn}, uses a hybrid structure of long short-term memory (LSTM) and CNN. We increase the kernel size from 3 to 32 to enhance the receptive field and better adapt to our dataset. The third AF detector is built using CNN, LSTM, and an attention mechanism \cite{bahdanau_attention}, following the Attentional Convolutional Long Short-Term Memory Neural Network~(AC-LSTM) proposed by Jin et al. \cite{model_ac_lstm}, with the same kernel size adjustment.

Our ResNet-based detector achieves the highest F1 score of 97.0\%, outperforming the CNN-based model by 11.8\%, the LSTM-CNN model by 8.9\%, and the AC-LSTM model by 6.5\%. This result demonstrates the effectiveness of the ResNet-based detector in accurately detecting AF.

\subsection{System Robustness}

\subsubsection{Impact of Background Noise}
\label{s: evaluation.robustness.noise}

\begin{figure}
  \centering
  \includegraphics[width=4.5in]{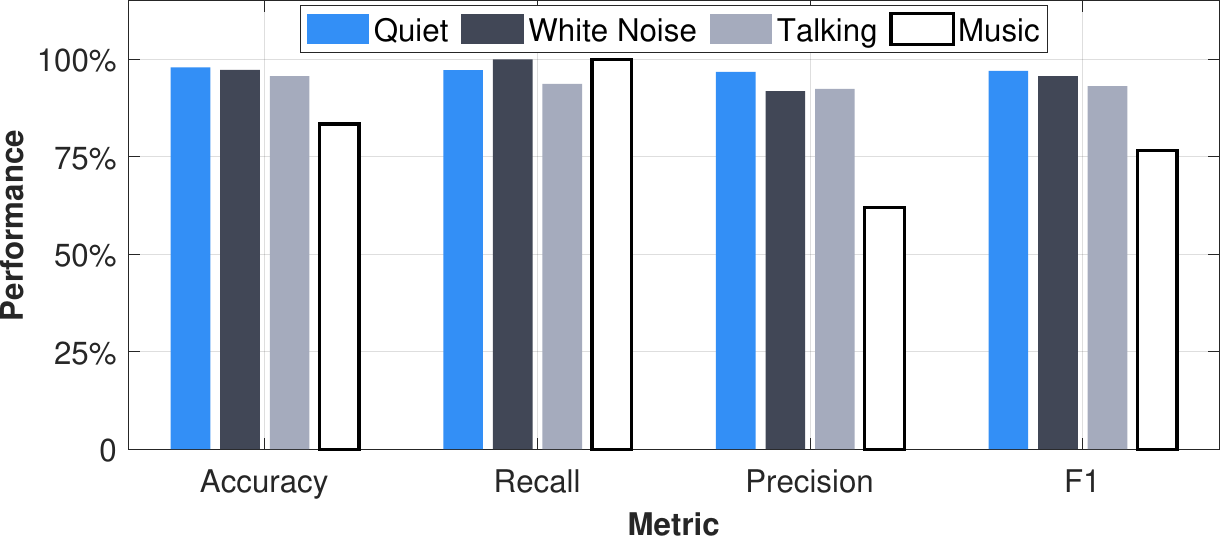}
  \caption{System performance comparison in quiet environments and three different noisy environments.}
  \label{f: robustness.noise}
\end{figure}

In our prior studies, all data are collected in quiet environments without significant interference. To evaluate the robustness of our system, we test it in three different noisy environments: \textbf{(a) White Noise:} a smartphone is placed 30 cm away from the experimental smartphone and plays white noise at a volume over 50 dB. \textbf{(b) Talking:} the participant engages in conversation with others at a volume over 50 dB. \textbf{(c) Music:} a smartphone is placed 30 cm away from the experimental smartphone and plays random pop music at a volume over 50 dB. For each noisy environment setting, an extended dataset is collected from 5 subjects (music and talking) or 4 subjects (white noise), with one subject being an AF subject and the remaining subjects being non-AF subjects. The AF detector is trained on the dataset discussed in Sec.~\ref{s: evaluation.overall_performance}, with all overlapping subjects in the extended dataset removed, and is then tested on the extended dataset.

Fig.~\ref{f: robustness.noise} presents the test results. The system maintains relatively high performance in white noise and talking environments, with F1 scores of 95.7\% and 93.1\%, respectively. However, system performance significantly decreases in the music environment, with an F1 score of 76.5\%. This decline may be due to the unexpected high-frequency components in the random pop music played during the experiments, which interfere with the acquisition of pulse waves. Additionally, the recall in white noise and music environments is higher than in quiet environments, which may be attributed to the difference in the AF subject ratio across the different datasets.

\subsubsection{Performance on Different Devices}
\begin{figure}[ht]
  \centering
  \begin{minipage}[b]{0.48\linewidth}
    \centering
    \includegraphics[width=\linewidth]{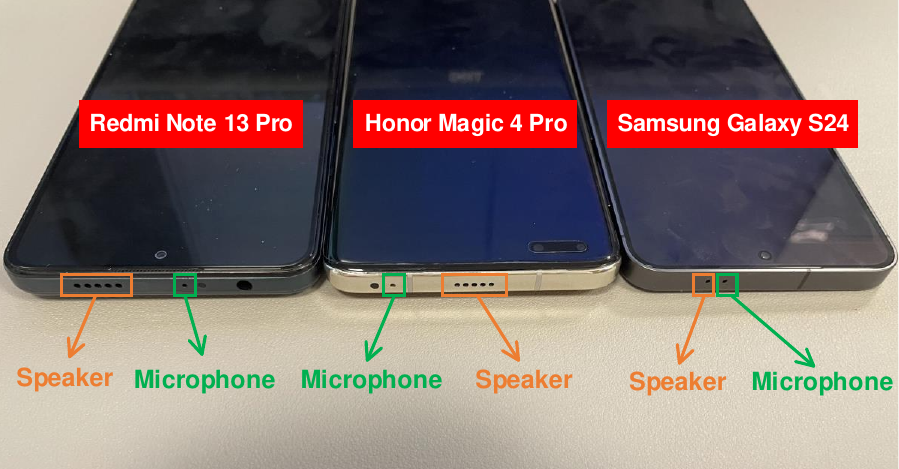}
    \caption{Top microphone and speaker arrangements of Redmi 13 Pro, Honor Magic 4 Pro, and Samsung Galaxy S24.}
    \label{f: robustness.device.mic_speaker_arrangement}
  \end{minipage}
  \hfill 
  \begin{minipage}[b]{0.48\linewidth}
    \centering
    \includegraphics[width=\linewidth]{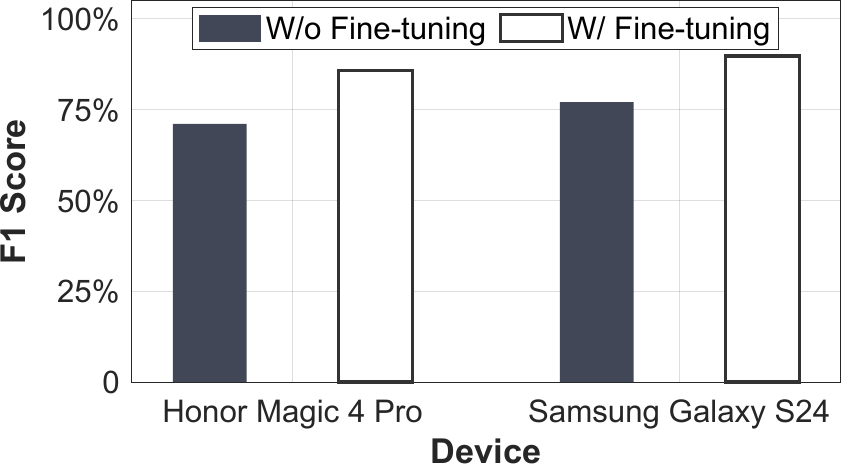}
    \caption{System performance on different devices with fine-tuning and without fine-tuning.}
    \label{f: robustness.device.performance}
  \end{minipage}
\end{figure}
In the previous experiments, raw data are collected using the Redmi Note 13 Pro. To evaluate the generalizability of our system across different commercially off-the-shelf (COTS) smartphones, additional experiments are conducted using the Honor Magic 4 Pro and Samsung Galaxy S24. These devices feature distinct and representative microphone and speaker layouts. Fig.~\ref{f: robustness.device.mic_speaker_arrangement} illustrates the microphone and speaker layouts of the three smartphones. The padding around the microphone and speaker is relatively wide for both the Redmi Note 13 Pro and Honor Magic 4 Pro, whereas it is narrower for the Samsung Galaxy S24. Additionally, the speaker on the Redmi Note 13 Pro is positioned closer to the edge of the device, while the microphone on the Honor Magic 4 Pro is located near the edge. For each new device, an extended test dataset is collected with two AF subjects and three non-AF subjects. The AF detector is trained on the dataset discussed in Sec.~\ref{s: evaluation.overall_performance}, with all overlapping subjects in the extended dataset removed, and is then tested on the extended dataset.

The comparison of the primary system performance (without fine-tuning) on the three devices is depicted in Fig.~\ref{f: robustness.device.performance}. We observe relatively low system performance on the two new devices. The F1 scores of the Honor Magic 4 Pro and Samsung Galaxy S24 are 71.1\% and 77.1\%, which are much lower than the F1 score of the Redmi Note 13 Pro, which is 97.0\%. The poor performance may be attributed to the AF detector mistakenly capturing some device-specific channel properties. To verify this hypothesis, we fine-tune the AF detector with one subject randomly selected from the extended dataset and validate the system on the remaining data. The performance of the fine-tuned system is shown in Fig.~\ref{f: robustness.device.performance}. The figure shows a significant performance improvement, with the F1 scores of the two devices increasing to 85.7\% and 89.8\%, confirming our hypothesis. This suggests that our system can be adapted to other devices and maintain high performance with a small amount of data for fine-tuning.

%% file: Body/related.tex
\section{Related Work}
\label{s: related}
In this section, we review the related works and divide them into two categories: AF detection and acoustic sensing for vital signs. 
\subsection{AF Detection}
In the field of AF detection, three primary techniques for acquiring cardiac signals prevail: ECG, PPG, and seismocardiography (SCG). These methods yield signals containing several features crucial for AF diagnosis, including RMSSD in the time domain, power in the frequency domain, and sample entropy in the entropy domain~\cite{ML1}. Traditional approaches typically employ statistical analysis of these features, utilizing thresholds to identify AF events. Meanwhile, machine learning strategies have become increasingly popular, leveraging a variety of features to train models for AF classification.

\subsubsection{ECG-based Approaches}
The electrocardiogram (ECG) is an important diagnostic tool for detecting various cardiac diseases, including atrial fibrillation, by measuring the electrical activity of the heart. Although this method is accurate, its complex setup and high cost make daily monitoring at home impractical, and the interpretation of ECGs requires specialized knowledge, increasing the burden of use on the user. To address these challenges, various smartphone-based ECG systems~\cite{ECG2, ECG3, MobileECG-1, MobileECG-2, MobileECG-3, MobileECG-4, MobileECG-5, MobileECG-6} have been developed, such as the AliveCor iPhone ECG and wrist-worn devices such as the Apple Watch, which simplify testing and embed atrial fibrillation detection algorithms~\cite{ECG2, ECG3}. While convenient, these innovations tend to be costly and still require additional sensors, limiting their widespread use, especially among older adults at higher risk for AF. Unlike the aforementioned research work, our approach utilizes the built-in microphone and speaker of a commercial smartphone, bypassing the need for additional sensors and making it easier to use and more cost-effective.

\subsubsection{PPG-based Approaches}

For mobile AF detection systems, previous studies have primarily relied on photoplethysmography (PPG), a non-invasive method that measures blood volume changes by illuminating the skin to detect pulse waves. Various techniques have been proposed to distinguish AF from NSR by analyzing features of the PPG waveform. Bashar \etal~\cite{WristPPG-1} identifies heart rate irregularity as a key indicator of AF. The system proposed in~\cite{PPG2} identifies AF utilizing statistical analysis. Since PPG waveform peaks are considered analogous to R waves in ECG waveforms, many methods have attempted to extract features from PPG signals for input into various machine learning models~\cite{ML1, ML2, ML3, ML4, ML5, ML6}. Although PPG-based approaches are widely used for AF detection, long-time wearing of PPG sensors may cause skin burning. Additionally, recent studies~\cite{PPG3, PPGSkinTone} have shown that PPG signal quality can be affected by factors such as skin tone and level of obesity, limiting the applicability of PPG methods. In contrast, our approach leverages the ubiquitous smartphones, offering a more accessible and convenient solution for AF detection without the constraints associated with PPG-based techniques. By utilizing the existing hardware of smartphones, our method circumvents the challenges posed by individual factors, ensuring both reliability and effectiveness across a wide range of users.

\subsubsection{SCG-based Approaches}
Recent studies have also proposed using SCG to measure HRV. SCG is a non-invasive technique that measures the cardiogenic acceleration of the chest. It utilizes an inertial measurement unit (IMU) to capture the motion induced by cardiac activity. Wang \etal ~\cite{SCG1} introduces a method to detect AF by using the accelerometer of mobile phones by recording small chest movements. However, this method does not achieve on-device analysis; instead, the reconstructed ECG signals are transmitted to physicians for real-time evaluation. In situations where a physician is unavailable, this approach does not meet the requirements for continuous monitoring. Additionally, transforming SCG signals to ECG signals may introduce noise, potentially leading to misinterpretations and erroneous medical decisions.  Koivisto \etal ~\cite{SCG2} collects SCG data using a prototype placed on subjects' chests while requiring them to lie down to align with a predefined coordinate system. The requirement for the user to lie down restricts its applicability. In contrast, our method directly detects AF by analyzing cardiac signals extracted from sound waves. This approach eliminates the potential noises introduced by signal translation and only requires the user to keep their wrist still for approximately 30 seconds during data collection, offering comfort and convenience for users.

\subsection{Acoustic Sensing For Vital Sign}
Recent advancements in acoustic sensing for monitoring vital signs have gained attention, leveraging the unique capabilities of acoustic signals for sensing. C. Wang \etal ~\cite{acous1} utilizes soft structural components with piezoelectric composites embedded to emit ultrasound of high frequency and record the signals corresponding to pulsating anterior and posterior walls to detect pulse waves so as to estimate blood pressure. F. Wang \etal ~\cite{acous2} also leverages piezoelectric transducer-embedded soft structural components but measures blood flow velocity through the Doppler effect. In addition, earphones can serve as devices for transmitting and receiving acoustic signals. Balaji \etal ~\cite{acous3} discovers that due to the heart's left-sided position, the time it takes for blood to reach the blood vessels in the left and right ears differs. Then, blood pressure can be estimated using this time difference in the received signals from earphones. Fan \etal ~\cite{acous4} observes that the subtle changes in the ear canal caused by blood vessel deformations would modulate ultrasound signals. Furthermore, COTS devices without special design also have the potential to obtain vital signs. Qian's research~\cite{acous5} indicates that it is possible to utilize the smartphone's speaker and microphone to capture acoustic signals modulated by chest motion, allowing for the extraction of heartbeat information from these signals. Wang \etal ~\cite{acous6} proposes a Correlation-based Frequency Modulated Continuous Wave method and implements a system for human respiration monitoring using commercial microphones and speakers. These existing works inspire and encourage us to develop an acoustic-based AF detection system deployed on COTS smartphones.

%% file: Body/discussion.tex
\section{Discussion}
\label{s: discussion}
 
Despite the effectiveness of \systemname in detecting AF, it has several limitations. This section outlines these limitations and suggests areas for future improvement.

\textbf{Variety of subjects.} In this study, we recruited 23 subjects from our university and the affiliated hospital for system evaluation. However, to better evaluate the robustness of \systemname, a more diverse subject pool from various social backgrounds is needed. We leave this for our future work.

\textbf{Interference from Other Arrhythmias.} \systemname may be affected by other arrhythmias, such as atrial flutter and premature contractions, which share similar features with AF. Future work should involve studying a diverse range of arrhythmias to enhance the system's differentiation capabilities.

\textbf{Privacy and Security.} Our system leverages mobile phones' speakers and microphones to probe users' pulse waves for AF detection. However, the use of microphones may compromise users' privacy and introduce vulnerabilities. Future work will explore ways to mitigate these risks and ensure the security and privacy of users' data.

\textbf{Influence of Device Position.} The acquisition of high-quality pulse waves heavily depends on the device's position. Although \systemname includes a \textit{Pulse Wave Quality Assessment} module to provide guidance on device position and filter out poor-quality data, it still requires some user efforts. Future work should explore more user-friendly approaches to probe pulse waves.

%% file: Body/conclusion.tex
\section{Conclusion}
\label{s: conclusion}
In conclusion, this paper presents \systemname, a system specifically developed for AF detection. \systemname  explores the potential of pulse wave acquisition from the wrist using smartphone speakers and microphones. We propose a robust and noise-resistant pulse wave probing method. In addition, the signal quality is enhanced by introducing a novel pulse wave purification pipeline. Then, A resnet-based AF detection model is proposed to achieve accurate and reliable AF detection. We have implemented a real-time analysis application that effectively collects and analyzes data. Through extensive experiments involving 9 AF patients and 14 individuals without AF, our system demonstrates a remarkable average accuracy of 97.9\%, recall of 97.2\%, precision of 96.8\%, specificity of 98.3\% and an impressive F1 score of 97.0\% in AF detection.

%% file: main.bbl

\begin{thebibliography}{65}


\ifx \showCODEN    \undefined \def \showCODEN     #1{\unskip}     \fi
\ifx \showDOI      \undefined \def \showDOI       #1{#1}\fi
\ifx \showISBNx    \undefined \def \showISBNx     #1{\unskip}     \fi
\ifx \showISBNxiii \undefined \def \showISBNxiii  #1{\unskip}     \fi
\ifx \showISSN     \undefined \def \showISSN      #1{\unskip}     \fi
\ifx \showLCCN     \undefined \def \showLCCN      #1{\unskip}     \fi
\ifx \shownote     \undefined \def \shownote      #1{#1}          \fi
\ifx \showarticletitle \undefined \def \showarticletitle #1{#1}   \fi
\ifx \showURL      \undefined \def \showURL       {\relax}        \fi
\providecommand\bibfield[2]{#2}
\providecommand\bibinfo[2]{#2}
\providecommand\natexlab[1]{#1}
\providecommand\showeprint[2][]{arXiv:#2}

\bibitem[Aliamiri and Shen(2018)]%
        {ML5}
\bibfield{author}{\bibinfo{person}{Alireza Aliamiri} {and} \bibinfo{person}{Yichen Shen}.} \bibinfo{year}{2018}\natexlab{}.
\newblock \showarticletitle{Deep learning based atrial fibrillation detection using wearable photoplethysmography sensor}. In \bibinfo{booktitle}{\emph{2018 IEEE EMBS International Conference on Biomedical \& Health Informatics (BHI)}}. IEEE, \bibinfo{pages}{442--445}.
\newblock


\bibitem[Bahdanau et~al\mbox{.}(2016)]%
        {bahdanau_attention}
\bibfield{author}{\bibinfo{person}{Dzmitry Bahdanau}, \bibinfo{person}{Kyunghyun Cho}, {and} \bibinfo{person}{Yoshua Bengio}.} \bibinfo{year}{2016}\natexlab{}.
\newblock \bibinfo{title}{Neural Machine Translation by Jointly Learning to Align and Translate}.
\newblock
\newblock
\showeprint[arxiv]{1409.0473}~[cs.CL]
\urldef\tempurl%
\url{https://arxiv.org/abs/1409.0473}
\showURL{%
\tempurl}


\bibitem[Balaji et~al\mbox{.}(2023)]%
        {acous3}
\bibfield{author}{\bibinfo{person}{Ananta~Narayanan Balaji}, \bibinfo{person}{Andrea Ferlini}, \bibinfo{person}{Fahim Kawsar}, {and} \bibinfo{person}{Alessandro Montanari}.} \bibinfo{year}{2023}\natexlab{}.
\newblock \showarticletitle{Stereo-bp: Non-invasive blood pressure sensing with earables}. In \bibinfo{booktitle}{\emph{Proceedings of the 24th International Workshop on Mobile Computing Systems and Applications}}. \bibinfo{pages}{96--102}.
\newblock


\bibitem[Bashar et~al\mbox{.}(2019a)]%
        {WristPPG-1}
\bibfield{author}{\bibinfo{person}{Syed~Khairul Bashar}, \bibinfo{person}{Dong Han}, \bibinfo{person}{Shirin Hajeb-Mohammadalipour}, \bibinfo{person}{Eric Ding}, \bibinfo{person}{Cody Whitcomb}, \bibinfo{person}{David~D. McManus}, {and} \bibinfo{person}{Ki~H. Chon}.} \bibinfo{year}{2019}\natexlab{a}.
\newblock \showarticletitle{Atrial Fibrillation Detection from Wrist Photoplethysmography Signals Using Smartwatches}.
\newblock \bibinfo{journal}{\emph{Scientific Reports}} \bibinfo{volume}{9}, \bibinfo{number}{1} (\bibinfo{date}{21 Oct} \bibinfo{year}{2019}), \bibinfo{pages}{15054}.
\newblock
\showISSN{2045-2322}
\urldef\tempurl%
\url{https://doi.org/10.1038/s41598-019-49092-2}
\showDOI{\tempurl}


\bibitem[Bashar et~al\mbox{.}(2019b)]%
        {WristPPGIMU-1}
\bibfield{author}{\bibinfo{person}{Syed~Khairul Bashar}, \bibinfo{person}{Dong Han}, \bibinfo{person}{Shirin Hajeb-Mohammadalipour}, \bibinfo{person}{Eric Ding}, \bibinfo{person}{Cody Whitcomb}, \bibinfo{person}{David~D. McManus}, {and} \bibinfo{person}{Ki~H. Chon}.} \bibinfo{year}{2019}\natexlab{b}.
\newblock \showarticletitle{Atrial {Fibrillation} {Detection} from {Wrist} {Photoplethysmography} {Signals} {Using} {Smartwatches}}.
\newblock \bibinfo{journal}{\emph{Scientific Reports}} \bibinfo{volume}{9}, \bibinfo{number}{1} (\bibinfo{date}{Oct.} \bibinfo{year}{2019}), \bibinfo{pages}{15054}.
\newblock
\showISSN{2045-2322}
\urldef\tempurl%
\url{https://doi.org/10.1038/s41598-019-49092-2}
\showDOI{\tempurl}
\newblock
\shownote{Publisher: Nature Publishing Group}.


\bibitem[Bashar et~al\mbox{.}(2018)]%
        {CameraPPG-8}
\bibfield{author}{\bibinfo{person}{Syed~Khairul Bashar}, \bibinfo{person}{Dong Han}, \bibinfo{person}{Apurv Soni}, \bibinfo{person}{David~D. McManus}, {and} \bibinfo{person}{Ki~H. Chon}.} \bibinfo{year}{2018}\natexlab{}.
\newblock \showarticletitle{Developing a novel noise artifact detection algorithm for smartphone {PPG} signals: {Preliminary} results}. In \bibinfo{booktitle}{\emph{2018 {IEEE} {EMBS} {International} {Conference} on {Biomedical} \& {Health} {Informatics} ({BHI})}}. \bibinfo{publisher}{IEEE}, \bibinfo{address}{Las Vegas, NV, USA}, \bibinfo{pages}{79--82}.
\newblock
\showISBNx{978-1-5386-2405-0}
\urldef\tempurl%
\url{https://doi.org/10.1109/BHI.2018.8333374}
\showDOI{\tempurl}


\bibitem[Bonomi et~al\mbox{.}(2016)]%
        {WristPPGIMU-2}
\bibfield{author}{\bibinfo{person}{Alberto Bonomi}, \bibinfo{person}{Fons Schipper}, \bibinfo{person}{Linda Eerikainen}, \bibinfo{person}{Jenny Margarito}, \bibinfo{person}{Ronald Aarts}, \bibinfo{person}{Saeed Babaeizadeh}, \bibinfo{person}{Helma De~Morree}, {and} \bibinfo{person}{Lukas Dekker}.} \bibinfo{year}{2016}\natexlab{}.
\newblock \showarticletitle{Atrial {Fibrillation} {Detection} {Using} {Photo}:plethysmography and {Acceleration} {Data} at the {Wrist}}.
\newblock
\urldef\tempurl%
\url{https://doi.org/10.22489/CinC.2016.081-339}
\showDOI{\tempurl}


\bibitem[Boonya-Ananta et~al\mbox{.}(2021)]%
        {PPG3}
\bibfield{author}{\bibinfo{person}{Tananant Boonya-Ananta}, \bibinfo{person}{Andres~J Rodriguez}, \bibinfo{person}{VN Du~Le}, \bibinfo{person}{Jessica~C Ramella-Roman}, {et~al\mbox{.}}} \bibinfo{year}{2021}\natexlab{}.
\newblock \showarticletitle{Monte Carlo analysis of optical heart rate sensors in commercial wearables: the effect of skin tone and obesity on the photoplethysmography (PPG) signal}.
\newblock \bibinfo{journal}{\emph{Biomedical optics express}} \bibinfo{volume}{12}, \bibinfo{number}{12} (\bibinfo{year}{2021}), \bibinfo{pages}{7445--7457}.
\newblock


\bibitem[Chan and Choy(2017)]%
        {MobileECG-4}
\bibfield{author}{\bibinfo{person}{Ngai-yin Chan} {and} \bibinfo{person}{Chi-chung Choy}.} \bibinfo{year}{2017}\natexlab{}.
\newblock \showarticletitle{Screening for atrial fibrillation in 13 122 {Hong} {Kong} citizens with smartphone electrocardiogram}.
\newblock \bibinfo{journal}{\emph{Heart}} \bibinfo{volume}{103}, \bibinfo{number}{1} (\bibinfo{date}{Jan.} \bibinfo{year}{2017}), \bibinfo{pages}{24--31}.
\newblock
\showISSN{1355-6037, 1468-201X}
\urldef\tempurl%
\url{https://doi.org/10.1136/heartjnl-2016-309993}
\showDOI{\tempurl}


\bibitem[Chan et~al\mbox{.}(2018)]%
        {MobileECG-2}
\bibfield{author}{\bibinfo{person}{Ngai-Yin Chan}, \bibinfo{person}{Chi-Chung Choy}, \bibinfo{person}{Chi-Kin Chan}, {and} \bibinfo{person}{Chung-Wah Siu}.} \bibinfo{year}{2018}\natexlab{}.
\newblock \showarticletitle{Effectiveness of a nongovernmental organization–led large-scale community atrial fibrillation screening program using the smartphone electrocardiogram: {An} observational cohort study}.
\newblock \bibinfo{journal}{\emph{Heart Rhythm}} \bibinfo{volume}{15}, \bibinfo{number}{9} (\bibinfo{date}{Sept.} \bibinfo{year}{2018}), \bibinfo{pages}{1306--1311}.
\newblock
\showISSN{15475271}
\urldef\tempurl%
\url{https://doi.org/10.1016/j.hrthm.2018.06.006}
\showDOI{\tempurl}


\bibitem[Chong et~al\mbox{.}(2018)]%
        {CameraPPG-6}
\bibfield{author}{\bibinfo{person}{Jo~Woon Chong}, \bibinfo{person}{Chae~Ho Cho}, \bibinfo{person}{Fatemehsadat Tabei}, \bibinfo{person}{Duy Le-Anh}, \bibinfo{person}{Nada Esa}, \bibinfo{person}{David~D. Mcmanus}, {and} \bibinfo{person}{Ki~H. Chon}.} \bibinfo{year}{2018}\natexlab{}.
\newblock \showarticletitle{Motion and {Noise} {Artifact}-{Resilient} {Atrial} {Fibrillation} {Detection} {Using} a {Smartphone}}.
\newblock \bibinfo{journal}{\emph{IEEE Journal on Emerging and Selected Topics in Circuits and Systems}} \bibinfo{volume}{8}, \bibinfo{number}{2} (\bibinfo{date}{June} \bibinfo{year}{2018}), \bibinfo{pages}{230--239}.
\newblock
\showISSN{2156-3357, 2156-3365}
\urldef\tempurl%
\url{https://doi.org/10.1109/JETCAS.2018.2818185}
\showDOI{\tempurl}


\bibitem[Conroy et~al\mbox{.}(2017)]%
        {PPG2}
\bibfield{author}{\bibinfo{person}{Thomas Conroy}, \bibinfo{person}{Jairo~Hernandez Guzman}, \bibinfo{person}{Burr Hall}, \bibinfo{person}{Gill Tsouri}, {and} \bibinfo{person}{Jean-Philippe Couderc}.} \bibinfo{year}{2017}\natexlab{}.
\newblock \showarticletitle{Detection of atrial fibrillation using an earlobe photoplethysmographic sensor}.
\newblock \bibinfo{journal}{\emph{Physiological measurement}} \bibinfo{volume}{38}, \bibinfo{number}{10} (\bibinfo{year}{2017}), \bibinfo{pages}{1906}.
\newblock


\bibitem[Corino et~al\mbox{.}(2017)]%
        {WristPPGIMU-3}
\bibfield{author}{\bibinfo{person}{Valentina D~A Corino}, \bibinfo{person}{Rita Laureanti}, \bibinfo{person}{Lorenzo Ferranti}, \bibinfo{person}{Giorgio Scarpini}, \bibinfo{person}{Federico Lombardi}, {and} \bibinfo{person}{Luca~T Mainardi}.} \bibinfo{year}{2017}\natexlab{}.
\newblock \showarticletitle{Detection of atrial fibrillation episodes using a wristband device}.
\newblock \bibinfo{journal}{\emph{Physiological Measurement}} \bibinfo{volume}{38}, \bibinfo{number}{5} (\bibinfo{date}{May} \bibinfo{year}{2017}), \bibinfo{pages}{787--799}.
\newblock
\showISSN{0967-3334, 1361-6579}
\urldef\tempurl%
\url{https://doi.org/10.1088/1361-6579/aa5dd7}
\showDOI{\tempurl}


\bibitem[Das et~al\mbox{.}(2022)]%
        {ML6}
\bibfield{author}{\bibinfo{person}{Sarkar Snigdha~Sarathi Das}, \bibinfo{person}{Subangkar~Karmaker Shanto}, \bibinfo{person}{Masum Rahman}, \bibinfo{person}{Md~Saiful Islam}, \bibinfo{person}{Atif~Hasan Rahman}, \bibinfo{person}{Mohammad~M Masud}, {and} \bibinfo{person}{Mohammed~Eunus Ali}.} \bibinfo{year}{2022}\natexlab{}.
\newblock \showarticletitle{BayesBeat: Reliable atrial fibrillation detection from noisy photoplethysmography data}.
\newblock \bibinfo{journal}{\emph{Proceedings of the ACM on Interactive, Mobile, Wearable and Ubiquitous Technologies}} \bibinfo{volume}{6}, \bibinfo{number}{1} (\bibinfo{year}{2022}), \bibinfo{pages}{1--21}.
\newblock


\bibitem[Ebrahimzadeh et~al\mbox{.}(2018)]%
        {ML3}
\bibfield{author}{\bibinfo{person}{Elias Ebrahimzadeh}, \bibinfo{person}{Maede Kalantari}, \bibinfo{person}{Mohammadamin Joulani}, \bibinfo{person}{Reza~Shahrokhi Shahraki}, \bibinfo{person}{Farahnaz Fayaz}, {and} \bibinfo{person}{Fereshteh Ahmadi}.} \bibinfo{year}{2018}\natexlab{}.
\newblock \showarticletitle{Prediction of paroxysmal Atrial Fibrillation: A machine learning based approach using combined feature vector and mixture of expert classification on HRV signal}.
\newblock \bibinfo{journal}{\emph{Computer methods and programs in biomedicine}}  \bibinfo{volume}{165} (\bibinfo{year}{2018}), \bibinfo{pages}{53--67}.
\newblock


\bibitem[Eerikäinen et~al\mbox{.}(2018)]%
        {WristPPG-5}
\bibfield{author}{\bibinfo{person}{Linda~M Eerikäinen}, \bibinfo{person}{Alberto~G Bonomi}, \bibinfo{person}{Fons Schipper}, \bibinfo{person}{Lukas R~C Dekker}, \bibinfo{person}{Rik Vullings}, \bibinfo{person}{Helma~M de Morree}, {and} \bibinfo{person}{Ronald~M Aarts}.} \bibinfo{year}{2018}\natexlab{}.
\newblock \showarticletitle{Comparison between electrocardiogram- and photoplethysmogram-derived features for atrial fibrillation detection in free-living conditions}.
\newblock \bibinfo{journal}{\emph{Physiological Measurement}} \bibinfo{volume}{39}, \bibinfo{number}{8} (\bibinfo{date}{aug} \bibinfo{year}{2018}), \bibinfo{pages}{084001}.
\newblock
\urldef\tempurl%
\url{https://doi.org/10.1088/1361-6579/aad2c0}
\showDOI{\tempurl}


\bibitem[Fallet et~al\mbox{.}(2019)]%
        {WristPPGIMU-5}
\bibfield{author}{\bibinfo{person}{Sibylle Fallet}, \bibinfo{person}{Mathieu Lemay}, \bibinfo{person}{Philippe Renevey}, \bibinfo{person}{Célestin Leupi}, \bibinfo{person}{Etienne Pruvot}, {and} \bibinfo{person}{Jean-Marc Vesin}.} \bibinfo{year}{2019}\natexlab{}.
\newblock \showarticletitle{Can one detect atrial fibrillation using a wrist-type photoplethysmographic device?}
\newblock \bibinfo{journal}{\emph{Medical \& Biological Engineering \& Computing}} \bibinfo{volume}{57}, \bibinfo{number}{2} (\bibinfo{date}{Feb.} \bibinfo{year}{2019}), \bibinfo{pages}{477--487}.
\newblock
\showISSN{1741-0444}
\urldef\tempurl%
\url{https://doi.org/10.1007/s11517-018-1886-0}
\showDOI{\tempurl}


\bibitem[Fan et~al\mbox{.}(2023)]%
        {acous4}
\bibfield{author}{\bibinfo{person}{Xiaoran Fan}, \bibinfo{person}{David Pearl}, \bibinfo{person}{Richard Howard}, \bibinfo{person}{Longfei Shangguan}, {and} \bibinfo{person}{Trausti Thormundsson}.} \bibinfo{year}{2023}\natexlab{}.
\newblock \showarticletitle{APG: Audioplethysmography for Cardiac Monitoring in Hearables}. In \bibinfo{booktitle}{\emph{Proceedings of the 29th Annual International Conference on Mobile Computing and Networking}}. \bibinfo{pages}{1--15}.
\newblock


\bibitem[Fox et~al\mbox{.}(2007)]%
        {resting_heartrate}
\bibfield{author}{\bibinfo{person}{Kim Fox}, \bibinfo{person}{Jeffrey~S. Borer}, \bibinfo{person}{A.~John Camm}, \bibinfo{person}{Nicolas Danchin}, \bibinfo{person}{Roberto Ferrari}, \bibinfo{person}{Jose L.~Lopez Sendon}, \bibinfo{person}{Philippe~Gabriel Steg}, \bibinfo{person}{Jean-Claude Tardif}, \bibinfo{person}{Luigi Tavazzi}, \bibinfo{person}{Michal Tendera}, {and} \bibinfo{person}{null null}.} \bibinfo{year}{2007}\natexlab{}.
\newblock \showarticletitle{Resting Heart Rate in Cardiovascular Disease}.
\newblock \bibinfo{journal}{\emph{Journal of the American College of Cardiology}} \bibinfo{volume}{50}, \bibinfo{number}{9} (\bibinfo{year}{2007}), \bibinfo{pages}{823--830}.
\newblock
\urldef\tempurl%
\url{https://doi.org/10.1016/j.jacc.2007.04.079}
\showDOI{\tempurl}
\showeprint{https://www.jacc.org/doi/pdf/10.1016/j.jacc.2007.04.079}


\bibitem[Guo et~al\mbox{.}(2021a)]%
        {waltz}
\bibfield{author}{\bibinfo{person}{Junchen Guo}, \bibinfo{person}{Meng Jin}, \bibinfo{person}{Yuan He}, \bibinfo{person}{Weiguo Wang}, {and} \bibinfo{person}{Yunhao Liu}.} \bibinfo{year}{2021}\natexlab{a}.
\newblock \showarticletitle{Dancing Waltz with Ghosts: Measuring Sub-mm-Level 2D Rotor Orbit with a Single mmWave Radar}. In \bibinfo{booktitle}{\emph{Proceedings of the 20th International Conference on Information Processing in Sensor Networks (Co-Located with CPS-IoT Week 2021)}} (Nashville, TN, USA) \emph{(\bibinfo{series}{IPSN '21})}. \bibinfo{publisher}{Association for Computing Machinery}, \bibinfo{address}{New York, NY, USA}, \bibinfo{pages}{77–92}.
\newblock
\showISBNx{9781450380980}
\urldef\tempurl%
\url{https://doi.org/10.1145/3412382.3458258}
\showDOI{\tempurl}


\bibitem[Guo et~al\mbox{.}(2021b)]%
        {ML2}
\bibfield{author}{\bibinfo{person}{Yutao Guo}, \bibinfo{person}{Hao Wang}, \bibinfo{person}{Hui Zhang}, \bibinfo{person}{Tong Liu}, \bibinfo{person}{Luping Li}, \bibinfo{person}{Lingjie Liu}, \bibinfo{person}{Maolin Chen}, \bibinfo{person}{Yundai Chen}, {and} \bibinfo{person}{Gregory~YH Lip}.} \bibinfo{year}{2021}\natexlab{b}.
\newblock \showarticletitle{Photoplethysmography-based machine learning approaches for atrial fibrillation prediction: a report from the huawei heart study}.
\newblock \bibinfo{journal}{\emph{JACC: Asia}} \bibinfo{volume}{1}, \bibinfo{number}{3} (\bibinfo{year}{2021}), \bibinfo{pages}{399--408}.
\newblock


\bibitem[Haberman et~al\mbox{.}(2015)]%
        {MobileECG-5}
\bibfield{author}{\bibinfo{person}{Zachary~C. Haberman}, \bibinfo{person}{Ryan~T. Jahn}, \bibinfo{person}{Rupan Bose}, \bibinfo{person}{Han Tun}, \bibinfo{person}{Jerold~S. Shinbane}, \bibinfo{person}{Rahul~N. Doshi}, \bibinfo{person}{Philip~M. Chang}, {and} \bibinfo{person}{Leslie~A. Saxon}.} \bibinfo{year}{2015}\natexlab{}.
\newblock \showarticletitle{Wireless {Smartphone} {ECG} {Enables} {Large}-{Scale} {Screening} in {Diverse} {Populations}}.
\newblock \bibinfo{journal}{\emph{Journal of Cardiovascular Electrophysiology}} \bibinfo{volume}{26}, \bibinfo{number}{5} (\bibinfo{year}{2015}), \bibinfo{pages}{520--526}.
\newblock
\showISSN{1540-8167}
\urldef\tempurl%
\url{https://doi.org/10.1111/jce.12634}
\showDOI{\tempurl}
\newblock
\shownote{\_eprint: https://onlinelibrary.wiley.com/doi/pdf/10.1111/jce.12634}.


\bibitem[Han et~al\mbox{.}(2020a)]%
        {WristPPGIMU-4}
\bibfield{author}{\bibinfo{person}{Dong Han}, \bibinfo{person}{Syed~Khairul Bashar}, \bibinfo{person}{Fahimeh Mohagheghian}, \bibinfo{person}{Eric Ding}, \bibinfo{person}{Cody Whitcomb}, \bibinfo{person}{David~D. McManus}, {and} \bibinfo{person}{Ki~H. Chon}.} \bibinfo{year}{2020}\natexlab{a}.
\newblock \showarticletitle{Premature {Atrial} and {Ventricular} {Contraction} {Detection} {Using} {Photoplethysmographic} {Data} from a {Smartwatch}}.
\newblock \bibinfo{journal}{\emph{Sensors}} \bibinfo{volume}{20}, \bibinfo{number}{19} (\bibinfo{date}{Oct.} \bibinfo{year}{2020}), \bibinfo{pages}{5683}.
\newblock
\showISSN{1424-8220}
\urldef\tempurl%
\url{https://doi.org/10.3390/s20195683}
\showDOI{\tempurl}


\bibitem[Han et~al\mbox{.}(2020b)]%
        {MobileECG-6}
\bibfield{author}{\bibinfo{person}{Dong Han}, \bibinfo{person}{Syed~Khairul Bashar}, \bibinfo{person}{Fahimeh Mohagheghian}, \bibinfo{person}{Eric Ding}, \bibinfo{person}{Cody Whitcomb}, \bibinfo{person}{David~D. McManus}, {and} \bibinfo{person}{Ki~H. Chon}.} \bibinfo{year}{2020}\natexlab{b}.
\newblock \showarticletitle{Premature {Atrial} and {Ventricular} {Contraction} {Detection} {Using} {Photoplethysmographic} {Data} from a {Smartwatch}}.
\newblock \bibinfo{journal}{\emph{Sensors}} \bibinfo{volume}{20}, \bibinfo{number}{19} (\bibinfo{date}{Oct.} \bibinfo{year}{2020}), \bibinfo{pages}{5683}.
\newblock
\showISSN{1424-8220}
\urldef\tempurl%
\url{https://doi.org/10.3390/s20195683}
\showDOI{\tempurl}


\bibitem[He et~al\mbox{.}(2015)]%
        {he2015deepresiduallearningimage}
\bibfield{author}{\bibinfo{person}{Kaiming He}, \bibinfo{person}{Xiangyu Zhang}, \bibinfo{person}{Shaoqing Ren}, {and} \bibinfo{person}{Jian Sun}.} \bibinfo{year}{2015}\natexlab{}.
\newblock \bibinfo{title}{Deep Residual Learning for Image Recognition}.
\newblock
\newblock
\showeprint[arxiv]{1512.03385}~[cs.CV]
\urldef\tempurl%
\url{https://arxiv.org/abs/1512.03385}
\showURL{%
\tempurl}


\bibitem[Hong et~al\mbox{.}(2020)]%
        {hong_resnet1d}
\bibfield{author}{\bibinfo{person}{Shenda Hong}, \bibinfo{person}{Yanbo Xu}, \bibinfo{person}{Alind Khare}, \bibinfo{person}{Satria Priambada}, \bibinfo{person}{Kevin Maher}, \bibinfo{person}{Alaa Aljiffry}, \bibinfo{person}{Jimeng Sun}, {and} \bibinfo{person}{Alexey Tumanov}.} \bibinfo{year}{2020}\natexlab{}.
\newblock \showarticletitle{HOLMES: Health OnLine Model Ensemble Serving for Deep Learning Models in Intensive Care Units}. In \bibinfo{booktitle}{\emph{Proceedings of the 26th ACM SIGKDD International Conference on Knowledge Discovery \& Data Mining}}. \bibinfo{pages}{1614--1624}.
\newblock


\bibitem[Isakadze and Martin(2020)]%
        {ECG3}
\bibfield{author}{\bibinfo{person}{Nino Isakadze} {and} \bibinfo{person}{Seth~S Martin}.} \bibinfo{year}{2020}\natexlab{}.
\newblock \showarticletitle{How useful is the smartwatch ECG?}
\newblock \bibinfo{journal}{\emph{Trends in cardiovascular medicine}} \bibinfo{volume}{30}, \bibinfo{number}{7} (\bibinfo{year}{2020}), \bibinfo{pages}{442--448}.
\newblock


\bibitem[Jiang et~al\mbox{.}(2020)]%
        {mmVib}
\bibfield{author}{\bibinfo{person}{Chengkun Jiang}, \bibinfo{person}{Junchen Guo}, \bibinfo{person}{Yuan He}, \bibinfo{person}{Meng Jin}, \bibinfo{person}{Shuai Li}, {and} \bibinfo{person}{Yunhao Liu}.} \bibinfo{year}{2020}\natexlab{}.
\newblock \showarticletitle{mmVib: micrometer-level vibration measurement with mmwave radar}. In \bibinfo{booktitle}{\emph{Proceedings of the 26th Annual International Conference on Mobile Computing and Networking}} (London, United Kingdom) \emph{(\bibinfo{series}{MobiCom '20})}. \bibinfo{publisher}{Association for Computing Machinery}, \bibinfo{address}{New York, NY, USA}, Article \bibinfo{articleno}{45}, \bibinfo{numpages}{13}~pages.
\newblock
\showISBNx{9781450370851}
\urldef\tempurl%
\url{https://doi.org/10.1145/3372224.3419202}
\showDOI{\tempurl}


\bibitem[Jin et~al\mbox{.}(2020)]%
        {model_ac_lstm}
\bibfield{author}{\bibinfo{person}{Yanrui Jin}, \bibinfo{person}{Chengjin Qin}, \bibinfo{person}{Yixiang Huang}, \bibinfo{person}{Wenyi Zhao}, {and} \bibinfo{person}{Chengliang Liu}.} \bibinfo{year}{2020}\natexlab{}.
\newblock \showarticletitle{Multi-domain modeling of atrial fibrillation detection with twin attentional convolutional long short-term memory neural networks}.
\newblock \bibinfo{journal}{\emph{Knowledge-Based Systems}}  \bibinfo{volume}{193} (\bibinfo{year}{2020}), \bibinfo{pages}{105460}.
\newblock
\showISSN{0950-7051}
\urldef\tempurl%
\url{https://doi.org/10.1016/j.knosys.2019.105460}
\showDOI{\tempurl}


\bibitem[{Jinseok Lee} et~al\mbox{.}(2013)]%
        {CameraPPG-2}
\bibfield{author}{\bibinfo{person}{{Jinseok Lee}}, \bibinfo{person}{{Yunyoung Nam}}, \bibinfo{person}{David~D. McManus}, {and} \bibinfo{person}{Ki~H. Chon}.} \bibinfo{year}{2013}\natexlab{}.
\newblock \showarticletitle{Time-{Varying} {Coherence} {Function} for {Atrial} {Fibrillation} {Detection}}.
\newblock \bibinfo{journal}{\emph{IEEE Transactions on Biomedical Engineering}} \bibinfo{volume}{60}, \bibinfo{number}{10} (\bibinfo{date}{Oct.} \bibinfo{year}{2013}), \bibinfo{pages}{2783--2793}.
\newblock
\showISSN{0018-9294, 1558-2531}
\urldef\tempurl%
\url{https://doi.org/10.1109/TBME.2013.2264721}
\showDOI{\tempurl}


\bibitem[Koivisto et~al\mbox{.}(2015)]%
        {SCG2}
\bibfield{author}{\bibinfo{person}{Tero Koivisto}, \bibinfo{person}{Mikko P{\"a}nk{\"a}{\"a}l{\"a}}, \bibinfo{person}{Tero Hurnanen}, \bibinfo{person}{Tuija Vasankari}, \bibinfo{person}{Tuomas Kiviniemi}, \bibinfo{person}{Antti Saraste}, {and} \bibinfo{person}{Juhani Airaksinen}.} \bibinfo{year}{2015}\natexlab{}.
\newblock \showarticletitle{Automatic detection of atrial fibrillation using MEMS accelerometer}. In \bibinfo{booktitle}{\emph{2015 Computing in Cardiology Conference (CinC)}}. IEEE, \bibinfo{pages}{829--832}.
\newblock


\bibitem[Kornej et~al\mbox{.}(2020)]%
        {AF_significance}
\bibfield{author}{\bibinfo{person}{Jelena Kornej}, \bibinfo{person}{Christin~S. Börschel}, \bibinfo{person}{Emelia~J. Benjamin}, {and} \bibinfo{person}{Renate~B. Schnabel}.} \bibinfo{year}{2020}\natexlab{}.
\newblock \showarticletitle{Epidemiology of Atrial Fibrillation in the 21st Century}.
\newblock \bibinfo{journal}{\emph{Circulation Research}} \bibinfo{volume}{127}, \bibinfo{number}{1} (\bibinfo{year}{2020}), \bibinfo{pages}{4--20}.
\newblock
\urldef\tempurl%
\url{https://doi.org/10.1161/CIRCRESAHA.120.316340}
\showDOI{\tempurl}
\showeprint{https://www.ahajournals.org/doi/pdf/10.1161/CIRCRESAHA.120.316340}


\bibitem[Krivoshei et~al\mbox{.}(2016)]%
        {CameraPPG-3}
\bibfield{author}{\bibinfo{person}{Lian Krivoshei}, \bibinfo{person}{Stefan Weber}, \bibinfo{person}{Thilo Burkard}, \bibinfo{person}{Anna Maseli}, \bibinfo{person}{Noe Brasier}, \bibinfo{person}{Michael Kühne}, \bibinfo{person}{David Conen}, \bibinfo{person}{Thomas Huebner}, \bibinfo{person}{Andrea Seeck}, {and} \bibinfo{person}{Jens Eckstein}.} \bibinfo{year}{2016}\natexlab{}.
\newblock \showarticletitle{Smart detection of atrial fibrillation}.
\newblock \bibinfo{journal}{\emph{Europace}} (\bibinfo{date}{July} \bibinfo{year}{2016}), \bibinfo{pages}{euw125}.
\newblock
\showISSN{1099-5129, 1532-2092}
\urldef\tempurl%
\url{https://doi.org/10.1093/europace/euw125}
\showDOI{\tempurl}


\bibitem[Lahdenoja et~al\mbox{.}(2017)]%
        {SCG3}
\bibfield{author}{\bibinfo{person}{Olli Lahdenoja}, \bibinfo{person}{Tero Hurnanen}, \bibinfo{person}{Zuhair Iftikhar}, \bibinfo{person}{Sami Nieminen}, \bibinfo{person}{Timo Knuutila}, \bibinfo{person}{Antti Saraste}, \bibinfo{person}{Tuomas Kiviniemi}, \bibinfo{person}{Tuija Vasankari}, \bibinfo{person}{Juhani Airaksinen}, \bibinfo{person}{Mikko P{\"a}nk{\"a}{\"a}l{\"a}}, {et~al\mbox{.}}} \bibinfo{year}{2017}\natexlab{}.
\newblock \showarticletitle{Atrial fibrillation detection via accelerometer and gyroscope of a smartphone}.
\newblock \bibinfo{journal}{\emph{IEEE Journal of Biomedical and Health Informatics}} \bibinfo{volume}{22}, \bibinfo{number}{1} (\bibinfo{year}{2017}), \bibinfo{pages}{108--118}.
\newblock


\bibitem[Lai et~al\mbox{.}(2020)]%
        {MobileECG-1}
\bibfield{author}{\bibinfo{person}{Dakun Lai}, \bibinfo{person}{Yuxiang Bu}, \bibinfo{person}{Ye Su}, \bibinfo{person}{Xinshu Zhang}, {and} \bibinfo{person}{Chang-Sheng Ma}.} \bibinfo{year}{2020}\natexlab{}.
\newblock \showarticletitle{A {Flexible} {Multilayered} {Dry} {Electrode} and {Assembly} to {Single}-{Lead} {ECG} {Patch} to {Monitor} {Atrial} {Fibrillation} in a {Real}-{Life} {Scenario}}.
\newblock \bibinfo{journal}{\emph{IEEE Sensors Journal}} \bibinfo{volume}{20}, \bibinfo{number}{20} (\bibinfo{date}{Oct.} \bibinfo{year}{2020}), \bibinfo{pages}{12295--12306}.
\newblock
\showISSN{1530-437X, 1558-1748, 2379-9153}
\urldef\tempurl%
\url{https://doi.org/10.1109/JSEN.2020.2999101}
\showDOI{\tempurl}


\bibitem[Lau et~al\mbox{.}(2013)]%
        {ECG2}
\bibfield{author}{\bibinfo{person}{Jerrett~K Lau}, \bibinfo{person}{Nicole Lowres}, \bibinfo{person}{Lis Neubeck}, \bibinfo{person}{David~B Brieger}, \bibinfo{person}{Raymond~W Sy}, \bibinfo{person}{Connor~D Galloway}, \bibinfo{person}{David~E Albert}, {and} \bibinfo{person}{Saul~B Freedman}.} \bibinfo{year}{2013}\natexlab{}.
\newblock \showarticletitle{iPhone ECG application for community screening to detect silent atrial fibrillation: a novel technology to prevent stroke}.
\newblock \bibinfo{journal}{\emph{International journal of cardiology}} \bibinfo{volume}{165}, \bibinfo{number}{1} (\bibinfo{year}{2013}), \bibinfo{pages}{193--194}.
\newblock


\bibitem[Lemay et~al\mbox{.}(2016)]%
        {WristPPG-2}
\bibfield{author}{\bibinfo{person}{Mathieu Lemay}, \bibinfo{person}{Sibylle Fallet}, \bibinfo{person}{Philippe Renevey}, \bibinfo{person}{Josep Solà}, \bibinfo{person}{Célestin Leupi}, \bibinfo{person}{Etienne Pruvot}, {and} \bibinfo{person}{Jean-Marc Vesin}.} \bibinfo{year}{2016}\natexlab{}.
\newblock \showarticletitle{Wrist-located optical device for atrial fibrillation screening: A clinical study on twenty patients}. In \bibinfo{booktitle}{\emph{2016 Computing in Cardiology Conference (CinC)}}. \bibinfo{pages}{681--684}.
\newblock
\showISSN{2325-887X}


\bibitem[Lippi et~al\mbox{.}(2020)]%
        {AF-Intro}
\bibfield{author}{\bibinfo{person}{Giuseppe Lippi}, \bibinfo{person}{Fabian Sanchis-Gomar}, {and} \bibinfo{person}{Gianfranco Cervellin}.} \bibinfo{year}{2020}\natexlab{}.
\newblock \showarticletitle{Global epidemiology of atrial fibrillation: An increasing epidemic and public health challenge}.
\newblock \bibinfo{journal}{\emph{Int J Stroke}} \bibinfo{volume}{16}, \bibinfo{number}{2} (\bibinfo{date}{Jan.} \bibinfo{year}{2020}), \bibinfo{pages}{217--221}.
\newblock


\bibitem[Marini et~al\mbox{.}(2005)]%
        {AF-With-Stroke}
\bibfield{author}{\bibinfo{person}{Carmine Marini}, \bibinfo{person}{Federica~De Santis}, \bibinfo{person}{Simona Sacco}, \bibinfo{person}{Tommasina Russo}, \bibinfo{person}{Luigi Olivieri}, \bibinfo{person}{Rocco Totaro}, {and} \bibinfo{person}{Antonio Carolei}.} \bibinfo{year}{2005}\natexlab{}.
\newblock \showarticletitle{Contribution of Atrial Fibrillation to Incidence and Outcome of Ischemic Stroke}.
\newblock \bibinfo{journal}{\emph{Stroke}} \bibinfo{volume}{36}, \bibinfo{number}{6} (\bibinfo{year}{2005}), \bibinfo{pages}{1115--1119}.
\newblock
\urldef\tempurl%
\url{https://doi.org/10.1161/01.STR.0000166053.83476.4a}
\showDOI{\tempurl}


\bibitem[McMANUS et~al\mbox{.}(2016)]%
        {CameraPPG-4}
\bibfield{author}{\bibinfo{person}{David~D. McMANUS}, \bibinfo{person}{Jo~Woon Chong}, \bibinfo{person}{Apurv Soni}, \bibinfo{person}{Jane~S. Saczynski}, \bibinfo{person}{Nada Esa}, \bibinfo{person}{Craig Napolitano}, \bibinfo{person}{Chad~E. Darling}, \bibinfo{person}{Edward Boyer}, \bibinfo{person}{Rochelle~K. Rosen}, \bibinfo{person}{Kevin~C. Floyd}, {and} \bibinfo{person}{Ki~H. Chon}.} \bibinfo{year}{2016}\natexlab{}.
\newblock \showarticletitle{{PULSE}-{SMART}: {Pulse}-{Based} {Arrhythmia} {Discrimination} {Using} a {Novel} {Smartphone} {Application}}.
\newblock \bibinfo{journal}{\emph{Journal of Cardiovascular Electrophysiology}} \bibinfo{volume}{27}, \bibinfo{number}{1} (\bibinfo{year}{2016}), \bibinfo{pages}{51--57}.
\newblock
\showISSN{1540-8167}
\urldef\tempurl%
\url{https://doi.org/10.1111/jce.12842}
\showDOI{\tempurl}
\newblock
\shownote{\_eprint: https://onlinelibrary.wiley.com/doi/pdf/10.1111/jce.12842}.


\bibitem[McManus et~al\mbox{.}(2013)]%
        {CameraPPG-1}
\bibfield{author}{\bibinfo{person}{David~D. McManus}, \bibinfo{person}{Jinseok Lee}, \bibinfo{person}{Oscar Maitas}, \bibinfo{person}{Nada Esa}, \bibinfo{person}{Rahul Pidikiti}, \bibinfo{person}{Alex Carlucci}, \bibinfo{person}{Josephine Harrington}, \bibinfo{person}{Eric Mick}, {and} \bibinfo{person}{Ki~H. Chon}.} \bibinfo{year}{2013}\natexlab{}.
\newblock \showarticletitle{A novel application for the detection of an irregular pulse using an iPhone 4S in patients with atrial fibrillation}.
\newblock \bibinfo{journal}{\emph{Heart Rhythm}} \bibinfo{volume}{10}, \bibinfo{number}{3} (\bibinfo{year}{2013}), \bibinfo{pages}{315--319}.
\newblock
\showISSN{1547-5271}
\urldef\tempurl%
\url{https://doi.org/10.1016/j.hrthm.2012.12.001}
\showDOI{\tempurl}


\bibitem[Nemati et~al\mbox{.}(2016)]%
        {WristPPG-4}
\bibfield{author}{\bibinfo{person}{Shamim Nemati}, \bibinfo{person}{Mohammad~M. Ghassemi}, \bibinfo{person}{Vaidehi Ambai}, \bibinfo{person}{Nino Isakadze}, \bibinfo{person}{Oleksiy Levantsevych}, \bibinfo{person}{Amit Shah}, {and} \bibinfo{person}{Gari~D. Clifford}.} \bibinfo{year}{2016}\natexlab{}.
\newblock \showarticletitle{Monitoring and detecting atrial fibrillation using wearable technology}. In \bibinfo{booktitle}{\emph{2016 38th Annual International Conference of the IEEE Engineering in Medicine and Biology Society (EMBC)}}. \bibinfo{pages}{3394--3397}.
\newblock
\showISSN{1558-4615}
\urldef\tempurl%
\url{https://doi.org/10.1109/EMBC.2016.7591456}
\showDOI{\tempurl}


\bibitem[Nguyen et~al\mbox{.}(2022)]%
        {ML4}
\bibfield{author}{\bibinfo{person}{Duc~Huy Nguyen}, \bibinfo{person}{Paul C-P Chao}, \bibinfo{person}{Chih-Chieh Chung}, \bibinfo{person}{Ray-Hua Horng}, {and} \bibinfo{person}{Bhaskar Choubey}.} \bibinfo{year}{2022}\natexlab{}.
\newblock \showarticletitle{Detecting Atrial Fibrillation in Real Time Based on PPG via Two CNNs for Quality Assessment and Detection}.
\newblock \bibinfo{journal}{\emph{IEEE Sensors Journal}} \bibinfo{volume}{22}, \bibinfo{number}{24} (\bibinfo{year}{2022}), \bibinfo{pages}{24102--24111}.
\newblock


\bibitem[Nurmaini et~al\mbox{.}(2020)]%
        {model_cnn}
\bibfield{author}{\bibinfo{person}{Siti Nurmaini}, \bibinfo{person}{Alexander Edo Tondas}, \bibinfo{person}{Annisa Darmawahyuni}, \bibinfo{person}{Muhammad~Naufal Rachmatullah}, \bibinfo{person}{Radiyati {Umi Partan}}, \bibinfo{person}{Firdaus Firdaus}, \bibinfo{person}{Bambang Tutuko}, \bibinfo{person}{Ferlita Pratiwi}, \bibinfo{person}{Andre~Herviant Juliano}, {and} \bibinfo{person}{Rahmi Khoirani}.} \bibinfo{year}{2020}\natexlab{}.
\newblock \showarticletitle{Robust detection of atrial fibrillation from short-term electrocardiogram using convolutional neural networks}.
\newblock \bibinfo{journal}{\emph{Future Generation Computer Systems}}  \bibinfo{volume}{113} (\bibinfo{year}{2020}), \bibinfo{pages}{304--317}.
\newblock
\showISSN{0167-739X}
\urldef\tempurl%
\url{https://doi.org/10.1016/j.future.2020.07.021}
\showDOI{\tempurl}


\bibitem[Orchard et~al\mbox{.}(2016)]%
        {MobileECG-3}
\bibfield{author}{\bibinfo{person}{Jessica Orchard}, \bibinfo{person}{Nicole Lowres}, \bibinfo{person}{S~Ben Freedman}, \bibinfo{person}{Laila Ladak}, \bibinfo{person}{William Lee}, \bibinfo{person}{Nicholas Zwar}, \bibinfo{person}{David Peiris}, \bibinfo{person}{Yasith Kamaladasa}, \bibinfo{person}{Jialin Li}, {and} \bibinfo{person}{Lis Neubeck}.} \bibinfo{year}{2016}\natexlab{}.
\newblock \showarticletitle{Screening for atrial fibrillation during influenza vaccinations by primary care nurses using a smartphone electrocardiograph ({iECG}): {A} feasibility study}.
\newblock \bibinfo{journal}{\emph{European Journal of Preventive Cardiology}} \bibinfo{volume}{23}, \bibinfo{number}{2\_suppl} (\bibinfo{date}{Oct.} \bibinfo{year}{2016}), \bibinfo{pages}{13--20}.
\newblock
\showISSN{2047-4873, 2047-4881}
\urldef\tempurl%
\url{https://doi.org/10.1177/2047487316670255}
\showDOI{\tempurl}


\bibitem[Paszke et~al\mbox{.}(2019)]%
        {pytorch-library}
\bibfield{author}{\bibinfo{person}{Adam Paszke}, \bibinfo{person}{Sam Gross}, \bibinfo{person}{Francisco Massa}, \bibinfo{person}{Adam Lerer}, \bibinfo{person}{James Bradbury}, \bibinfo{person}{Gregory Chanan}, \bibinfo{person}{Trevor Killeen}, \bibinfo{person}{Zeming Lin}, \bibinfo{person}{Natalia Gimelshein}, \bibinfo{person}{Luca Antiga}, \bibinfo{person}{Alban Desmaison}, \bibinfo{person}{Andreas Köpf}, \bibinfo{person}{Edward Yang}, \bibinfo{person}{Zach DeVito}, \bibinfo{person}{Martin Raison}, \bibinfo{person}{Alykhan Tejani}, \bibinfo{person}{Sasank Chilamkurthy}, \bibinfo{person}{Benoit Steiner}, \bibinfo{person}{Lu Fang}, \bibinfo{person}{Junjie Bai}, {and} \bibinfo{person}{Soumith Chintala}.} \bibinfo{year}{2019}\natexlab{}.
\newblock \bibinfo{title}{PyTorch: An Imperative Style, High-Performance Deep Learning Library}.
\newblock
\newblock
\showeprint[arxiv]{1912.01703}~[cs.LG]
\urldef\tempurl%
\url{https://arxiv.org/abs/1912.01703}
\showURL{%
\tempurl}


\bibitem[Proesmans et~al\mbox{.}(2019)]%
        {CameraPPG-9}
\bibfield{author}{\bibinfo{person}{Tine Proesmans}, \bibinfo{person}{Christophe Mortelmans}, \bibinfo{person}{Ruth Van~Haelst}, \bibinfo{person}{Frederik Verbrugge}, \bibinfo{person}{Pieter Vandervoort}, {and} \bibinfo{person}{Bert Vaes}.} \bibinfo{year}{2019}\natexlab{}.
\newblock \showarticletitle{Mobile {Phone}–{Based} {Use} of the {Photoplethysmography} {Technique} to {Detect} {Atrial} {Fibrillation} in {Primary} {Care}: {Diagnostic} {Accuracy} {Study} of the {FibriCheck} {App}}.
\newblock \bibinfo{journal}{\emph{JMIR mHealth and uHealth}} \bibinfo{volume}{7}, \bibinfo{number}{3} (\bibinfo{date}{March} \bibinfo{year}{2019}), \bibinfo{pages}{e12284}.
\newblock
\showISSN{2291-5222}
\urldef\tempurl%
\url{https://doi.org/10.2196/12284}
\showDOI{\tempurl}


\bibitem[Puranen et~al\mbox{.}(2020)]%
        {PPGSkinTone}
\bibfield{author}{\bibinfo{person}{Antti Puranen}, \bibinfo{person}{Tuomas Halkola}, \bibinfo{person}{Ole Kirkeby}, {and} \bibinfo{person}{Antti Vehkaoja}.} \bibinfo{year}{2020}\natexlab{}.
\newblock \showarticletitle{Effect of skin tone and activity on the performance of wrist-worn optical beat-to-beat heart rate monitoring}. In \bibinfo{booktitle}{\emph{2020 IEEE SENSORS}}. \bibinfo{pages}{1--4}.
\newblock
\urldef\tempurl%
\url{https://doi.org/10.1109/SENSORS47125.2020.9278523}
\showDOI{\tempurl}


\bibitem[Pänkäälä et~al\mbox{.}(2016)]%
        {SCGAF}
\bibfield{author}{\bibinfo{person}{Mikko Pänkäälä}, \bibinfo{person}{Tero Koivisto}, \bibinfo{person}{Olli Lahdenoja}, \bibinfo{person}{Tuomas Kiviniemi}, \bibinfo{person}{Antti Saraste}, \bibinfo{person}{Tuija Vasankari}, {and} \bibinfo{person}{Juhani Airaksinen}.} \bibinfo{year}{2016}\natexlab{}.
\newblock \showarticletitle{Detection of atrial fibrillation with seismocardiography}. In \bibinfo{booktitle}{\emph{2016 38th Annual International Conference of the IEEE Engineering in Medicine and Biology Society (EMBC)}}. \bibinfo{pages}{4369--4374}.
\newblock
\urldef\tempurl%
\url{https://doi.org/10.1109/EMBC.2016.7591695}
\showDOI{\tempurl}


\bibitem[Qian et~al\mbox{.}(2018)]%
        {acous5}
\bibfield{author}{\bibinfo{person}{Kun Qian}, \bibinfo{person}{Chenshu Wu}, \bibinfo{person}{Fu Xiao}, \bibinfo{person}{Yue Zheng}, \bibinfo{person}{Yi Zhang}, \bibinfo{person}{Zheng Yang}, {and} \bibinfo{person}{Yunhao Liu}.} \bibinfo{year}{2018}\natexlab{}.
\newblock \showarticletitle{Acousticcardiogram: Monitoring heartbeats using acoustic signals on smart devices}. In \bibinfo{booktitle}{\emph{IEEE INFOCOM 2018-IEEE conference on computer communications}}. IEEE, \bibinfo{pages}{1574--1582}.
\newblock


\bibitem[Savelieva and Camm(2000)]%
        {silentAF}
\bibfield{author}{\bibinfo{person}{Irina Savelieva} {and} \bibinfo{person}{A.~John Camm}.} \bibinfo{year}{2000}\natexlab{}.
\newblock \showarticletitle{Clinical Relevance of Silent Atrial Fibrillation: Prevalence, Prognosis, Quality of Life, and Management}.
\newblock \bibinfo{journal}{\emph{Journal of Interventional Cardiac Electrophysiology}} \bibinfo{volume}{4}, \bibinfo{number}{2} (\bibinfo{date}{01 Jun} \bibinfo{year}{2000}), \bibinfo{pages}{369--382}.
\newblock
\showISSN{1572-8595}
\urldef\tempurl%
\url{https://doi.org/10.1023/A:1009823001707}
\showDOI{\tempurl}


\bibitem[Schäck et~al\mbox{.}(2017)]%
        {CameraPPG-7}
\bibfield{author}{\bibinfo{person}{Tim Schäck}, \bibinfo{person}{Yosef Safi~Harb}, \bibinfo{person}{Michael Muma}, {and} \bibinfo{person}{Abdelhak~M. Zoubir}.} \bibinfo{year}{2017}\natexlab{}.
\newblock \showarticletitle{Computationally efficient algorithm for photoplethysmography-based atrial fibrillation detection using smartphones}. In \bibinfo{booktitle}{\emph{2017 39th {Annual} {International} {Conference} of the {IEEE} {Engineering} in {Medicine} and {Biology} {Society} ({EMBC})}}. \bibinfo{pages}{104--108}.
\newblock
\urldef\tempurl%
\url{https://doi.org/10.1109/EMBC.2017.8036773}
\showDOI{\tempurl}
\newblock
\shownote{ISSN: 1558-4615}.


\bibitem[Shan et~al\mbox{.}(2016)]%
        {ML1}
\bibfield{author}{\bibinfo{person}{Shih-Ming Shan}, \bibinfo{person}{Sung-Chun Tang}, \bibinfo{person}{Pei-Wen Huang}, \bibinfo{person}{Yu-Min Lin}, \bibinfo{person}{Wei-Han Huang}, \bibinfo{person}{Dar-Ming Lai}, {and} \bibinfo{person}{An-Yeu~Andy Wu}.} \bibinfo{year}{2016}\natexlab{}.
\newblock \showarticletitle{Reliable PPG-based algorithm in atrial fibrillation detection}. In \bibinfo{booktitle}{\emph{2016 IEEE Biomedical Circuits and Systems Conference (BioCAS)}}. IEEE, \bibinfo{pages}{340--343}.
\newblock


\bibitem[Shashikumar et~al\mbox{.}(2017)]%
        {WristPPG-6}
\bibfield{author}{\bibinfo{person}{Supreeth~Prajwal Shashikumar}, \bibinfo{person}{Amit~J. Shah}, \bibinfo{person}{Qiao Li}, \bibinfo{person}{Gari~D. Clifford}, {and} \bibinfo{person}{Shamim Nemati}.} \bibinfo{year}{2017}\natexlab{}.
\newblock \showarticletitle{A deep learning approach to monitoring and detecting atrial fibrillation using wearable technology}. In \bibinfo{booktitle}{\emph{2017 IEEE EMBS International Conference on Biomedical \& Health Informatics (BHI)}}. \bibinfo{pages}{141--144}.
\newblock
\urldef\tempurl%
\url{https://doi.org/10.1109/BHI.2017.7897225}
\showDOI{\tempurl}


\bibitem[Soni et~al\mbox{.}(2017)]%
        {CameraPPG-5}
\bibfield{author}{\bibinfo{person}{Apurv Soni}, \bibinfo{person}{Sunil Karna}, \bibinfo{person}{Harshil Patel}, \bibinfo{person}{Nisha Fahey}, \bibinfo{person}{Shyamsundar Raithatha}, \bibinfo{person}{Anna Handorf}, \bibinfo{person}{John Bostrom}, \bibinfo{person}{Syed Bashar}, \bibinfo{person}{Kandarp Talati}, \bibinfo{person}{Ravi Shah}, \bibinfo{person}{Robert~J Goldberg}, \bibinfo{person}{Sunil Thanvi}, \bibinfo{person}{Ajay~Gajanan Phatak}, \bibinfo{person}{Jeroan~J Allison}, \bibinfo{person}{Ki Chon}, \bibinfo{person}{Somashekhar~Marutirao Nimbalkar}, {and} \bibinfo{person}{David~D McManus}.} \bibinfo{year}{2017}\natexlab{}.
\newblock \showarticletitle{Study protocol for \textit{{S}} martphone \textit{{M}} onitoring for \textit{{A}} trial fibrillation in \textit{{R}} eal- \textit{{T}} ime in {India} ({SMART}-{India}): a community-based screening and referral programme}.
\newblock \bibinfo{journal}{\emph{BMJ Open}} \bibinfo{volume}{7}, \bibinfo{number}{12} (\bibinfo{date}{Dec.} \bibinfo{year}{2017}), \bibinfo{pages}{e017668}.
\newblock
\showISSN{2044-6055, 2044-6055}
\urldef\tempurl%
\url{https://doi.org/10.1136/bmjopen-2017-017668}
\showDOI{\tempurl}


\bibitem[Tarniceriu et~al\mbox{.}(2018)]%
        {WristPPG-3}
\bibfield{author}{\bibinfo{person}{Adrian Tarniceriu}, \bibinfo{person}{Jarkko Harju}, \bibinfo{person}{Zeinab~Rezaei Yousefi}, \bibinfo{person}{Antti Vehkaoja}, \bibinfo{person}{Jakub Parak}, \bibinfo{person}{Arvi Yli-Hankala}, {and} \bibinfo{person}{Ilkka Korhonen}.} \bibinfo{year}{2018}\natexlab{}.
\newblock \showarticletitle{The Accuracy of Atrial Fibrillation Detection from Wrist Photoplethysmography. A Study on Post-Operative Patients}. In \bibinfo{booktitle}{\emph{2018 40th Annual International Conference of the IEEE Engineering in Medicine and Biology Society (EMBC)}}. \bibinfo{pages}{1--4}.
\newblock
\showISSN{1558-4615}
\urldef\tempurl%
\url{https://doi.org/10.1109/EMBC.2018.8513197}
\showDOI{\tempurl}


\bibitem[Wang et~al\mbox{.}(2018a)]%
        {acous1}
\bibfield{author}{\bibinfo{person}{Chonghe Wang}, \bibinfo{person}{Xiaoshi Li}, \bibinfo{person}{Hongjie Hu}, \bibinfo{person}{Lin Zhang}, \bibinfo{person}{Zhenlong Huang}, \bibinfo{person}{Muyang Lin}, \bibinfo{person}{Zhuorui Zhang}, \bibinfo{person}{Zhenan Yin}, \bibinfo{person}{Brady Huang}, \bibinfo{person}{Hua Gong}, {et~al\mbox{.}}} \bibinfo{year}{2018}\natexlab{a}.
\newblock \showarticletitle{Monitoring of the central blood pressure waveform via a conformal ultrasonic device}.
\newblock \bibinfo{journal}{\emph{Nature biomedical engineering}} \bibinfo{volume}{2}, \bibinfo{number}{9} (\bibinfo{year}{2018}), \bibinfo{pages}{687--695}.
\newblock


\bibitem[Wang et~al\mbox{.}(2021)]%
        {acous2}
\bibfield{author}{\bibinfo{person}{Fengle Wang}, \bibinfo{person}{Peng Jin}, \bibinfo{person}{Yunlu Feng}, \bibinfo{person}{Ji Fu}, \bibinfo{person}{Peng Wang}, \bibinfo{person}{Xin Liu}, \bibinfo{person}{Yingchao Zhang}, \bibinfo{person}{Yinji Ma}, \bibinfo{person}{Yingyun Yang}, \bibinfo{person}{Aiming Yang}, {et~al\mbox{.}}} \bibinfo{year}{2021}\natexlab{}.
\newblock \showarticletitle{Flexible Doppler ultrasound device for the monitoring of blood flow velocity}.
\newblock \bibinfo{journal}{\emph{Science advances}} \bibinfo{volume}{7}, \bibinfo{number}{44} (\bibinfo{year}{2021}), \bibinfo{pages}{eabi9283}.
\newblock


\bibitem[Wang et~al\mbox{.}(2023a)]%
        {SCG1}
\bibfield{author}{\bibinfo{person}{Lei Wang}, \bibinfo{person}{Xingwei Wang}, \bibinfo{person}{Dalin Zhang}, \bibinfo{person}{Xiaolei Ma}, \bibinfo{person}{Yong Zhang}, \bibinfo{person}{Haipeng Dai}, \bibinfo{person}{Chenren Xu}, \bibinfo{person}{Zhijun Li}, {and} \bibinfo{person}{Tao Gu}.} \bibinfo{year}{2023}\natexlab{a}.
\newblock \showarticletitle{Knowing Your Heart Condition Anytime: User-Independent ECG Measurement Using Commercial Mobile Phones}.
\newblock \bibinfo{journal}{\emph{Proceedings of the ACM on Interactive, Mobile, Wearable and Ubiquitous Technologies}} \bibinfo{volume}{7}, \bibinfo{number}{3} (\bibinfo{year}{2023}), \bibinfo{pages}{1--28}.
\newblock


\bibitem[Wang et~al\mbox{.}(2023b)]%
        {swt1}
\bibfield{author}{\bibinfo{person}{Lei Wang}, \bibinfo{person}{Xingwei Wang}, \bibinfo{person}{Dalin Zhang}, \bibinfo{person}{Xiaolei Ma}, \bibinfo{person}{Yong Zhang}, \bibinfo{person}{Haipeng Dai}, \bibinfo{person}{Chenren Xu}, \bibinfo{person}{Zhijun Li}, {and} \bibinfo{person}{Tao Gu}.} \bibinfo{year}{2023}\natexlab{b}.
\newblock \showarticletitle{Knowing Your Heart Condition Anytime: User-Independent ECG Measurement Using Commercial Mobile Phones}.
\newblock \bibinfo{journal}{\emph{Proceedings of the ACM on Interactive, Mobile, Wearable and Ubiquitous Technologies}} \bibinfo{volume}{7}, \bibinfo{number}{3} (\bibinfo{year}{2023}), \bibinfo{pages}{1--28}.
\newblock


\bibitem[Wang et~al\mbox{.}(2018b)]%
        {acous6}
\bibfield{author}{\bibinfo{person}{Tianben Wang}, \bibinfo{person}{Daqing Zhang}, \bibinfo{person}{Yuanqing Zheng}, \bibinfo{person}{Tao Gu}, \bibinfo{person}{Xingshe Zhou}, {and} \bibinfo{person}{Bernadette Dorizzi}.} \bibinfo{year}{2018}\natexlab{b}.
\newblock \showarticletitle{C-FMCW based contactless respiration detection using acoustic signal}.
\newblock \bibinfo{journal}{\emph{Proceedings of the ACM on Interactive, Mobile, Wearable and Ubiquitous Technologies}} \bibinfo{volume}{1}, \bibinfo{number}{4} (\bibinfo{year}{2018}), \bibinfo{pages}{1--20}.
\newblock


\bibitem[Wei and Zhang(2015)]%
        {mTrack}
\bibfield{author}{\bibinfo{person}{Teng Wei} {and} \bibinfo{person}{Xinyu Zhang}.} \bibinfo{year}{2015}\natexlab{}.
\newblock \showarticletitle{mTrack: High-Precision Passive Tracking Using Millimeter Wave Radios}. In \bibinfo{booktitle}{\emph{Proceedings of the 21st Annual International Conference on Mobile Computing and Networking}} (Paris, France) \emph{(\bibinfo{series}{MobiCom '15})}. \bibinfo{publisher}{Association for Computing Machinery}, \bibinfo{address}{New York, NY, USA}, \bibinfo{pages}{117–129}.
\newblock
\showISBNx{9781450336192}
\urldef\tempurl%
\url{https://doi.org/10.1145/2789168.2790113}
\showDOI{\tempurl}


\bibitem[Zhang et~al\mbox{.}(2021b)]%
        {EarlyDetectionAf}
\bibfield{author}{\bibinfo{person}{Hanbin Zhang}, \bibinfo{person}{Li Zhu}, \bibinfo{person}{Viswam Nathan}, \bibinfo{person}{Jilong Kuang}, \bibinfo{person}{Jacob Kim}, \bibinfo{person}{Jun~Alex Gao}, {and} \bibinfo{person}{Jeffrey Olgin}.} \bibinfo{year}{2021}\natexlab{b}.
\newblock \showarticletitle{Towards Early Detection and Burden Estimation of Atrial Fibrillation in an Ambulatory Free-living Environment}.
\newblock \bibinfo{journal}{\emph{Proc. ACM Interact. Mob. Wearable Ubiquitous Technol.}} \bibinfo{volume}{5}, \bibinfo{number}{2}, Article \bibinfo{articleno}{86} (\bibinfo{date}{jun} \bibinfo{year}{2021}), \bibinfo{numpages}{19}~pages.
\newblock
\urldef\tempurl%
\url{https://doi.org/10.1145/3463503}
\showDOI{\tempurl}


\bibitem[Zhang et~al\mbox{.}(2021a)]%
        {model_lstm_cnn}
\bibfield{author}{\bibinfo{person}{Xiangyu Zhang}, \bibinfo{person}{Jianqing Li}, \bibinfo{person}{Zhipeng Cai}, \bibinfo{person}{Li Zhang}, \bibinfo{person}{Zhenghua Chen}, {and} \bibinfo{person}{Chengyu Liu}.} \bibinfo{year}{2021}\natexlab{a}.
\newblock \showarticletitle{Over-fitting suppression training strategies for deep learning-based atrial fibrillation detection}.
\newblock \bibinfo{journal}{\emph{Medical {\&} Biological Engineering {\&} Computing}} \bibinfo{volume}{59}, \bibinfo{number}{1} (\bibinfo{date}{01 Jan} \bibinfo{year}{2021}), \bibinfo{pages}{165--173}.
\newblock
\showISSN{1741-0444}
\urldef\tempurl%
\url{https://doi.org/10.1007/s11517-020-02292-9}
\showDOI{\tempurl}


\bibitem[Zhao et~al\mbox{.}(2024)]%
        {mmArrhythmia}
\bibfield{author}{\bibinfo{person}{Langcheng Zhao}, \bibinfo{person}{Rui Lyu}, \bibinfo{person}{Qi Lin}, \bibinfo{person}{Anfu Zhou}, \bibinfo{person}{Huanhuan Zhang}, \bibinfo{person}{Huadong Ma}, \bibinfo{person}{Jingjia Wang}, \bibinfo{person}{Chunli Shao}, {and} \bibinfo{person}{Yida Tang}.} \bibinfo{year}{2024}\natexlab{}.
\newblock \showarticletitle{mmArrhythmia: Contactless Arrhythmia Detection via mmWave Sensing}.
\newblock \bibinfo{journal}{\emph{Proc. ACM Interact. Mob. Wearable Ubiquitous Technol.}} \bibinfo{volume}{8}, \bibinfo{number}{1}, Article \bibinfo{articleno}{30} (\bibinfo{date}{mar} \bibinfo{year}{2024}), \bibinfo{numpages}{25}~pages.
\newblock
\urldef\tempurl%
\url{https://doi.org/10.1145/3643549}
\showDOI{\tempurl}


\end{thebibliography}
